\documentclass{nature-pre}
\usepackage{times}
\usepackage{amsmath}
\usepackage{amssymb}
\usepackage{physics}	
\usepackage{comment}
\usepackage{multirow}
\usepackage{physics}
\usepackage{makecell}
\usepackage{graphicx}
\graphicspath{./}
\usepackage{placeins}
\usepackage{threeparttable}
\usepackage{float}

\usepackage{hyperref}
\hypersetup{
 bookmarks=true,		
 unicode=false,			
 pdftoolbar=true,		
 pdffitwindow=false,		
 pdfstartview={FitH},		
 pdfcreator={pdflatex},		
 pdfproducer={vim},		
 pdfnewwindow=true,		
 colorlinks=true,		
 linktoc=page,			
 linkcolor=blue,		
 citecolor=blue,		
 filecolor=blue,		
 urlcolor=blue			
}

\title{Accurate and convergent energetics of color centers by wavefunction theory}

\author{Zsolt Benedek$^{1,2,3}$, \'{A}d\'{a}m Ganyecz$^{1,2}$, Anton Pershin$^{1,4}$, Viktor Iv\'{a}dy$^{2,3,5*}$, Gergely Barcza$^{1,2,*}$}

\begin{document}
\maketitle

\begin{affiliations}
\item  {HUN-REN Wigner Research Centre for Physics, PO Box 49, H-1525, Budapest, Hungary}
\item {MTA–ELTE Lend\"{u}let "Momentum" NewQubit Research Group, Pázmány Péter, Sétány 1/A, 1117 Budapest, Hungary}
\item{Department of Physics, Chemistry and Biology, Link\"oping University, SE-581 83 Link\"oping, Sweden}
\item{Department of Atomic Physics, Institute of Physics, Budapest University of Technology and Economics,  M\H{u}egyetem rakpart 3., H-1111 Budapest, Hungary}
\item{Department of Physics of Complex Systems, Eötvös Loránd University, Egyetem tér 1-3, H-1053 Budapest, Hungary}
\item[*] email: barcza.gergely@wigner.hu, ivady.viktor@ttk.elte.hu
\end{affiliations}

\date{\today}

\newpage

\begin{abstract}
Ab initio description of point defects in semiconductors, characterized by in-gap states of significant multideterminant character, presents a longstanding theoretical challenge for density functional theory (DFT) methods. In this study, we devise a wavefunction theory (WFT) based ab initio methodology as a competing alternative approach. Specifically, we apply perturbation theory (NEVPT2 level) on top of a defect-localized many-body wavefunction (CASSCF level), which provides a balanced description of dynamic and static correlation effects, respectively. This quantum chemical methodology, exemplified for the NV$^-$ center in diamond in this study, is not only used for the calculation of energies and properties, but also for geometry optimization, performed for each electronic state individually. By relaxing  cluster models of increasing size and investigating convergence behavior, we  quantitatively reproduce (i) the full energy spectrum of NV$^-$ including the recently characterized high-energy states, (ii) the effect of Jahn-Teller distortion on measurable properties, (iii) the fine structure of ground and excited states, (iv) the pressure dependence of zero-phonon lines. Our findings showcase that applying conventional wave-function-based quantum chemistry on carefully crafted clusters can be a robust routine tool for discussing defect-state energetics. 
\end{abstract}

\newpage

\maketitle

\section*{\large Introduction}
\subsection*{The challenge: multiconfigurational point-defect states}
\label{sec:intro_defect}
Owing to their unique magneto-optical properties, point-like defects in crystals that act as individual color centers have risen to fame with the advance of quantum technologies. In the last decade, such  solid-state color centers have been applied as  high-resolution nano-sized sensors\cite{DegenRMP2017,Aslam2023} thanks to their sensitivity to external electromagnetic fields, strain, and temperature. Furthermore, a large variety of single-photon emitters\cite{Chunnilall2014,Aharonovich2016,Zhang2020} has been identified in defected solids by now, which is an integral component in quantum computation\cite{Aspuru-Guzik2012} and quantum secure communication\cite{Lo2014}. Moreover, paramagnetic defects that enable spin-selective decay pathways could be used to create  quantum bits\cite{Ladd:Nature2010,Awschalom2013}, controllable through the optically detected magnetic resonance (ODMR) technique\cite{Suter2020}. 

From a theoretical point of view,  point defects hosted in wide-bandgap semiconductors behave like atoms featuring localized states in a screening medium of the bulk electrons. Spin-qubit applications relying on ODMR processes necessitate a complete understanding of the magneto-optical properties of the defect centers where strongly correlated singlet many-body states can play a vital role\cite{Gali_2019}. Therefore, the proper models of these color centers require simultaneous high-level treatment of both static and dynamic correlation effects corresponding to the defect and the embedding solid respectively\cite{Ma_2021,Pfaffle_2021,Muechler_2022}.

In the first place, the numerical exploration of the crystalline structures with point defect\cite{Zhang2020} is typically performed using density functional theory (DFT) based methods\cite{Jones2015}. This approach enables the computation of many relevant properties of color centers, such as formation energies, charge transition levels, spin states, hyperfine tensors, zero phonon lines, and photoluminescence spectra, albeit with varying accuracy, as summarized in reviews Ref.[\citenum{Freysoldt_2014,Dreyer_2018,Gali_2019,Zhang2020}].
However, the widely applied DFT is an inherently single-determinant method for ground state calculations, and it has limitations in describing states of strongly multideterminantal nature (also referred to as "multireference" or "multiconfigurational" character in the literature)\cite{Cohen2012,Verma-2020}. Thus, despite the tremendous theoretical progress in recent decades in studying correlated electronic states with DFT\cite{Makkar2021}, the quantitative description of solid state color centers still poses challenges\cite{Doherty-2013,Gali-2023}.
 
To further improve the models of spin-active defects, there is a great need for the development of a universally applicable wave function theory (WFT) based protocol that can accurately handle multiconfigurational problems. In fact, the defect community has already begun exploring post-DFT and post-Hartree-Fock methods for this purpose. Without attempting to provide an exhaustive list, we mention several common methods, including time-dependent DFT\cite{Raghavachari_2002,Gali2011,Jin2022,Jin_2023},  variational DFT\cite{Ivanov_2023}, configuration interaction constrained random phase approximation (CI-cRPA)\cite{Bockstedte2018} and GW based approximations\cite{Ma2010,Choi2012} as post-DFT approaches, as well as complete active space self-consistent field (CASSCF)\cite{Zyubin2009,Bhandari2021},  multireference configuration interaction (MRCI)\cite{Zyubin2009}, Monte Carlo configuration interaction (MCCI, FCIQMC)\cite{Delaney2010,Chen_2023},  density matrix renormalization group (DMRG)\cite{barcza_dmrg_2021} and equation of motion coupled cluster theory (EOM-CC)\cite{Li_2022} as wavefunction-based (i.e. post-HF) protocols. In addition, various quantum embedding theories have also been suggested\cite{Ma2020,Ma_2021,Muechler_2022,Sheng_2022,Chen_2023,Haldar_2023}.
These methods have mainly been benchmarked on the negatively charged nitrogen-vacancy (NV$^-$) center in diamond, which is the most relevant and extensively studied optically active spin defect to date\cite{Gali_2019}. While the distinct models concluded a largely consistent overall picture of the NV$^-$ vertical electronic structure, satisfactory quantitative agreement between theory and experiment could only be achieved by non-conventional combinations of many-body Hamiltonian and DFT on supercells\cite{Choi2012, Ma2020,Haldar_2023} or by pure many-body WFT on small clusters of questionably relevant size\cite{Bhandari2021,li2024excitedstatedynamicsopticallydetected}. Furthermore, the proper description of magneto-optical properties, taking into account geometry relaxation effects that are important for a detailed understanding of a defect center, has not been fully addressed within high-level WFT approaches. 

\subsection*{A potential solution: convergent wavefunction theory}
\label{sec:intro_methods}
In this work, we take a step forward in the direction of routine WFT modeling of qubits 
by presenting a novel computational strategy based entirely on readily available quantum chemical wave-function approaches, which have already been implemented in the commonly used codes. First, we highlight the critical importance of properly representing the hosting crystalline solid within finite diamond cluster models passivated by hydrogen atoms. The appropriate model size is determined based on careful convergence tests. By applying the CASSCF approach on the defect orbitals, we relax each electronic state of interest individually. The corresponding CASSCF electronic structure of the equilibrium geometry is improved by the energy correction resulted from the  second-order $n$–electron valence state perturbation theory (NEVPT2), which incorporates the dynamic correlation  effects of the embedding environment. We demonstrate the potential of this methodology by a comprehensive modeling of the prototypical color center, NV$^-$ in diamond,  aligning closely with the most firmly established experimental observations in the research field.

Below, we briefly summarize  the features of the applied approaches, highlighting the potential advantages of the CASSCF-NEVPT2 method for studying color centers with highly correlated electronic states arising from the atomic-like defect orbitals.

\subsubsection*{\textit{Static correlation (CASSCF)}}
The complete active space self-consistent field (CASSCF) method\cite{Siegbahn_1980,Roos_1980,Siegbahn_1981} captures the full range of correlation effects within a specific set ("active space") of molecular orbitals (MOs). During the CASSCF calculations, a two-step cycle of the full configuration interaction (FCI) solution and orbital optimization (i.e. mixing active and external orbitals to minimize the energy) is repeated until reaching convergence. 

Thus this method is capable of providing a practically exact solution in case all orbitals with non-negligible correlation effects are included in the active space. Unfortunately, owing to the exponential scaling of the approach with respect to the number of active orbitals, only a handful of orbitals (up to 20) can be set to active without further approximations. The remaining orbitals are kept frozen on the Hartree--Fock level during the solution of the configuration interaction problem, and only influence the correlation energy through some orbital mixing during the orbital optimization procedure.

Overall, the method provides a highly accurate description of the static correlation (i.e. the mixing of electronic states of different orbital occupation patterns) but it fails to assess dynamic correlation effect of the frozen inactive (doubly occupied) and virtual (unoccupied) orbitals. The resulting CASSCF wave function is qualitatively correct, and hence it was used for geometry optimization and vibrational analysis. However, neglecting the dynamic correlation generally results in dramatic errors in energy differences (gaps) between different electronic states. 

\begin{figure*}[t]
\begin{center}
\includegraphics[width=0.9\textwidth]{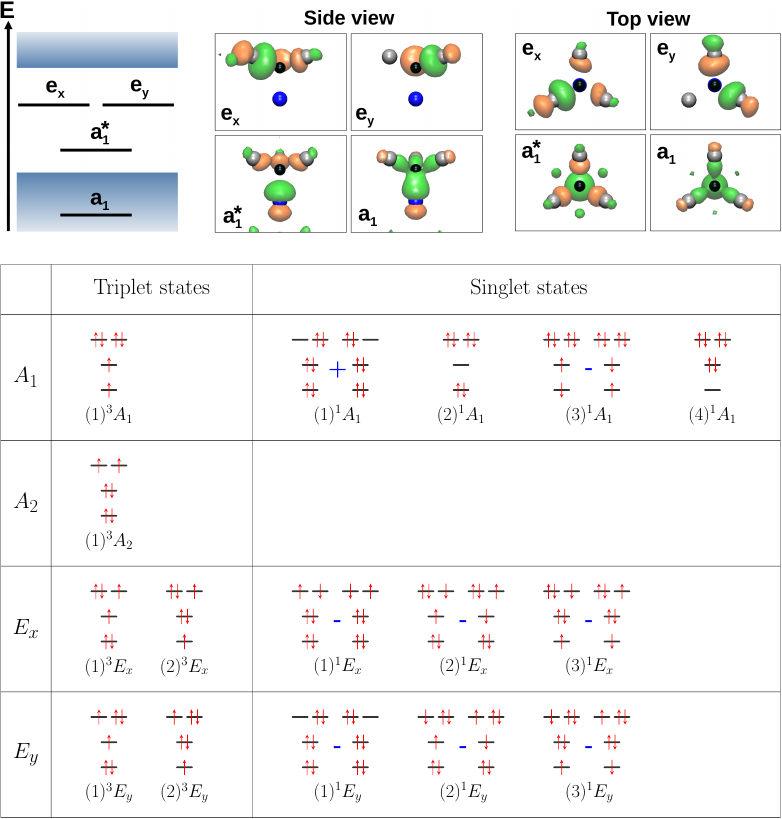}
\caption{Top: energy order and shape of NV$^-$ defect orbitals. In the schematic energy plot, the valence and conduction bands of diamond are visualized by blue. In the orbital plots, for the sake of visibility, only atoms  of the innermost shell around the vacancy are shown. Carbon and nitrogen atoms are distinguished by color gray and blue respectively. The vacancy site is highlighted in black. Bottom: Summary of  defect orbital configurations  used for the expansion of the CASSCF(6e,4o) many-body states. Configurations are tabulated corresponding to their $C_{3v}$ symmetry representation and spin. Within a given block, they are also labeled according to energy. 
}  
\label{fig:Configurations}
\end{center}
\end{figure*}

The proper choice of the active space for the CASSCF method is not always straightforward in computational chemistry\cite{Toth_2020}. Nevertheless, point defects in crystals typically yield a few of so-called defect orbitals, which define a chemically intuitive CAS. Note that these  defect-localized MOs lie inside and/or in the close vicinity of a large band gap  shaping the low-energy excitations. In the case of the NV center in diamond\cite{DohertyNVreview}, four relevant defect orbitals can be identified, specifically those that originate from the dangling bonds of the three carbon atoms and the nitrogen atom adjacent to the vacancy (Fig.~\ref{fig:Configurations}, left). These orbitals are altogether occupied by 6 electrons (accounting for 1 unpaired electron for each C atom, a lone electron pair for N and the local negative charge), which implies a CASSCF(6e,4o) procedure.

\subsubsection*{\textit{State-specific vs state-average CASSCF}}
The CASSCF approach is suitable for studying not only the ground state, but also the many-body excitations, as multiple solutions of the active-space FCI-problem (referred to as "roots" in CASSCF terminology) can be requested. The orbitals are optimized either for an ensemble of target electronic states in a state-averaged manner (SA-CASSCF), or for one selected state (root) in a state-specific manner (SS-CASSCF). 

In this work, both SS- and SA-CASSCF calculations were performed. When investigating a property which is peculiar to one well-defined electronic state (such as equilibrium geometry, vibrational analysis or zero-field splitting), we applied state-specific CASSCF to describe the state of interest as accurately as possible. On the other hand, for quantities that require multiple electronic states (such as excitation energies), a compromise must be found among the roots during the orbital optimization process, which can be achieved by state-averaging. 

We emphasize here that the number of electronic states involved in the state averaging is an additional parameter that influences the final CASSCF orbitals, hence the derived CASSCF-NEVPT2 results. The most simple approach is to include all studied electronic states (with equal weights) in all calculations. In this case, all energies become directly comparable (as the orbitals are optimized for the same set of roots); nevertheless, the more states are investigated in a single SA calculation, the less optimal the quality of the individual CASSCF wavefunctions becomes. Alternatively, to gain more accurate energy levels, one can perform two-state CASSCF-NEVPT2 calculations for each selected pair of electronic states and assemble the full energy spectrum based on Hess's law. In this work, both all-state and two-state averaging is applied, depending on the purpose of the calculation - see the Results section and the Supplementary Information (SI) for details.

\subsubsection*{\textit{ Dynamic correlation (NEVPT2)}}
If transitions between different electronic states are investigated, it is essential to take into account dynamic correlation on top of the (state-averaged) CASSCF wave function\cite{Schapiro2013}. Without this correction, theory might fail to reproduce excitation energies quantitatively\cite{Helmich2019}.

In the literature, various concepts have been developed to provide an \emph{a posteriori} correction to the CASSCF solution. Building on our previous work on point defects in hexagonal boron-nitride layers\cite{babar_quantum_2021,Benedek_2023}, we employ the second-order  $n$–electron valence state perturbation theory (NEVPT2)\cite{nevpt2,Angeli-2001a,Angeli2007}. This method, a type of multireference perturbative approach,is an extension of the second-order M\o ller-Plesset perturbation theory\cite{Moller_1934} to multireference systems. Owing to the construction of the zero-order  Dyall Hamiltonian\cite{Dyall_1995} and to the choice of perturbers, the NEVPT2 can provide a size-consistent theory on top of the CASSCF reference, that is free of intruder states and spin-contamination issues. Recent technical developments\cite{Guo_2021,Kollmar_2021,dlpno-nevpt2} have enabled the routine treatment of molecular systems with thousands of perturbing orbitals\cite{Guo_2023}.

\subsubsection*{\textit{ Fine structure and spin properties}}
The calculation of zero-field splitting (ZFS) parameters require spin-orbit coupled (SOC) and spin-spin coupled (SSC) states and their energy levels. In quantum chemistry, SOC and SSC coupling is introduced in a post-process manner: the required wave functions can be obtained in the basis of the CASSCF states  in the framework of quasi-degenerate perturbation theory (QDPT)\cite{Roemelt_2013}. The QDPT treatment assembles a SOC + SSC matrix is from the non-relativistic (Born-Oppenheimer, BO) CASSCF states (${\Psi_I}^{SM}$) as
\begin{equation*}
\mel{\Psi_I^{SM}}{\hat{H}_{BO}+\hat{H}_{SOC}+\hat{H}_{SSC}}{\Psi_J^{S'M'}} =  \delta_{IJ}\delta_{SS'}\delta_{MM'}E_I^S + \mel{\Psi_I^{SM}}{\hat{H}_{SOC}+\hat{H}_{SSC}}{\Psi_J^{S'M'}}  
\end{equation*}
where $I/J$, $S/S'$ and $M/M'$ indices refer to the number of the CASSCF root, its spin state and spin sublevel, respectively. The diagonal electronic energy, $E_I^S$, is taken into account at dynamic correlation corrected WFT level, herein NEVPT2. In our study, the $\langle \Psi_I^{SM}|{\hat{H}_{SOC}+\hat{H}_{SSC}}|\Psi_J^{S'M'}\rangle$ matrix elements are calculated between the previously obtained CASSCF roots using the spin-orbit mean-field (SOMF) approximation\cite{Neese_2005}. The spin sublevel dependence, which does not appear explicitly in non-relativistic CASSCF wavefunctions, can be introduced by means of Clebsch-Gordan coefficients, in accordance with the Wigner-Eckart theorem\cite{doi:Neese1998}.

The diagonalization of the QDPT matrix yields the coupled states as eigenvectors and the respective energy levels as eigenvalues, from which ZFS parameters can be extracted as energy differences.

Importantly, since CASSCF implementations work in restricted open-shell formalism (that is, the orbitals of $\alpha$ and $\beta$ channels are equal), the spin contamination issues of ZFS calculations, commonly observed for DFT\cite{biktagirov_spin_2020}, are avoided by construction.

\section*{\large Results}
\label{sec:results}
\bigbreak
\noindent {\bf Cluster model of the color center}
\label{sec:geom}
\begin{center}
\begin{figure*}[!t]
\includegraphics[width=\textwidth]{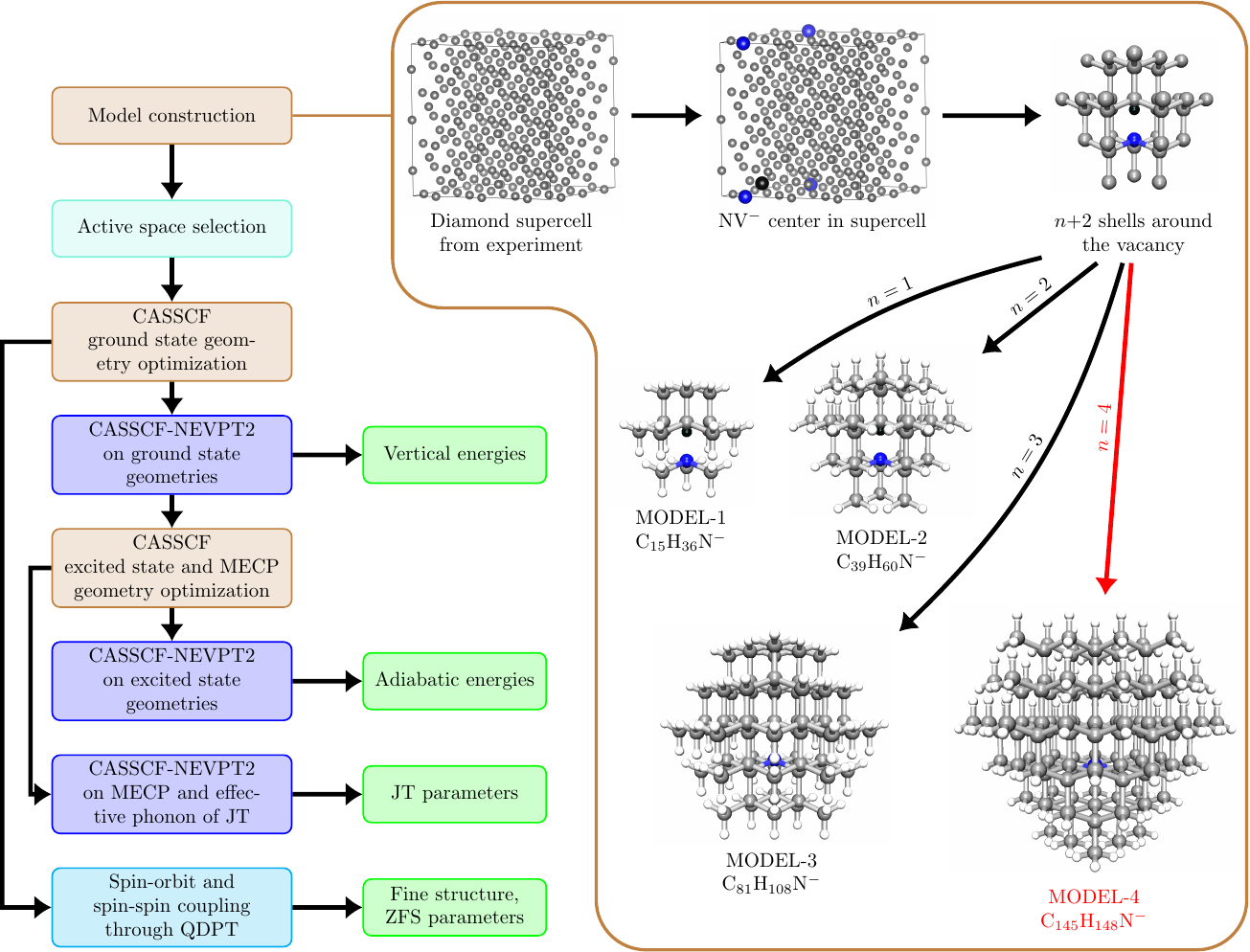}
\caption{Sketch of the applied computational protocol and  the NV molecular model (MODEL-$n$) construction. In the sketch,  steps related to geometry-construction/relaxation are in color brown, active space selection, CASSCF-NEVPT2 and CASSCF-QDPT calculations  highlighted in distinct shades of blue,  while the final results are in green.
 In the structures, C, N, and H atoms are denoted by color gray, blue and white respectively. The vacancy site is highlighted in black.  Recall, in MODEL-$n$, the inner $n+2$ coordination spheres around the vacancy were selected (as visualized for $n=1$ in the plot), carbons of shell $n+2$ were replaced by hydrogens to terminate the bonds as discussed in the main text. In the constrained geometry optimization,   only the inner $n$ coordination spheres were relaxed while the position of the carbons in shell $n+1$ and the terminal hydrogens in shell $n+2$ were kept fixed. MODEL-4, which is already large enough to provide a convergent model for the NV center, is highlighted in red.}
\label{fig:models}  
\end{figure*}
\end{center}
The NV defect in diamond comprises a substitutional nitrogen atom paired with an adjacent vacancy. In its negative charge state, it manifests a triplet ground state characterized by C$_{3v}$ symmetry\cite{Gali_2019}. In our molecular investigation, we employ quantum chemical models to simulate NV center embedded within nanodiamonds terminated with hydrogen at the surface. By progressively scaling up the cluster size, our objective is to accurately replicate the essential characteristics of the defected  bulk  crystal. 

Large-scale bulk calculations, e.g., Ref.[\citenum{Takacs_2024}],  indicate clearly that the defect center perturbs the perfect diamond crystal structure only in its close vicinity. Thus, to reflect the observed stiffness of the surrounding solid in our cluster models, we  optimized atomic positions only near the vacancy while enforcing the perfect diamond structure in the outer shells of the cluster.

In the following, we discuss this scheme, depicted in Fig.~\ref{fig:models}, in more details.
Initially, a sizeable pristine diamond of cubic crystal structure was formed with C-C bond distances, corresponding to the experimentally measured value of 1.54 \AA. 
In the supercell we implanted the NV defect by removing a carbon atom and replacing one of its adjacent carbons by nitrogen atom. The atoms were divided into "shells" according to their position relative to the vacancy: atoms at $n$ bond distance were assigned to the $n$th shell. 

In our most compact defected cluster model, denoted as MODEL-1, we consider a three-shelled structure, among which the outermost third shell is replaced by hydrogens to saturate all carbon valencies.  Motivated by hybrid theories\cite{Dapprich_1999} and by preserving the diamond crystal structure, we opt to adjust the length of outer C-H bonds  to the conventional value  of 1.09~\AA{}  without altering the positions of the third-shell H atoms (that were originally diamond C atoms) any further. During the  constrained geometry optimization, the position of the covering  hydrogens and the outer (second) carbon shell were  kept fixed  and we only allowed the relaxation of the inner atoms of the cluster. 

The general MODEL-$n$ can be constructed analogously: the innermost $n$ shells around the vacancy were optimized, by constraining the position of the carbon atoms in shell $n+1$ and the terminating hydrogens in shell $n+2$. In this work,  we studied $n$=1-4 cases, i.e., C$_{15}$H$_{36}$N$^-$, C$_{39}$H$_{60}$N$^-$, C$_{81}$H$_{108}$N$^-$, and C$_{145}$H$_{148}$N$^-$ clusters.

By constraining the position of the outer atoms, our model not only considers the compression of the surrounding diamond but it also  gives the chance to perform geometry optimization on a high-level theory for all states of interest due to the relatively limited number of atoms to be relaxed. While the accurate modeling of transition energies requires the comprehensive treatment of correlation effects, the state specific geometry optimization, which involves bond stretching processes, can be readily performed by capturing merely the static correlations. Therefore, we applied the CASSCF approach for the constrained relaxation of the cluster models. These calculations were feasible in practice since in MODEL-1, 2, 3, and 4 one only has to optimize the position of 4, 16, 40, and 82 atoms, respectively. Finally, we remark that the N-V axis is aligned with the magnetic axis in all calculations.

\bigbreak
\noindent {\bf Convergence with respect to model size and basis set}

The reliability of the applied CASSCF-NEVPT2 methodology depends on the size of the cluster model and the applied basis set. Sufficiently accurate description of the NV$^-$ center is only expected if the convergence with respect to the latter factors is demonstrated, i.e. the results negligibly deviate from those obtained from an infinitely large model using a complete basis set.

We began our investigations in this direction by performing a detailed analysis on basis dependence (see SI, section S2.1), investigating cc-pV$n$Z basis sets\cite{cc-pVDZ} with $n$= D, T, Q, 5 and 6 cardinal numbers. We found that the error of excitation energies arising from basis set truncation continuously decreases upon increasing the cluster model size, and cc-pVDZ practically recovers the basis set limit ($< 0.1$ eV error compared to cc-pVQZ) starting from MODEL-3. Therefore, the choice of the moderate cc-pVDZ basis set size is reasonable, given that the convergence of energies and properties with respect to model size can be obtained. In the following, we discuss the observed trends based on numerical data collected in sections S2-S5 of SI. 

First, we focus on the convergence behavior of excited-state energy levels with respect to model size (see also Tables S3 and S8 in SI). When increasing the size from MODEL-1 to MODEL-2, both the CASSCF energies and the NEVPT2 corrections show considerable alterations, which can even exceed 0.5 eV at the top of the energy spectrum. (Thus, MODEL-1 is only suitable for exploratory calculations.) In the next step, when changing to MODEL-3, the CASSCF energies remain practically unchanged (within 0.1 eV relative to MODEL-2 results for all states), but the dynamic correlation at NEVPT2 level has significant effect, exceeding 0.5 eV for the highest-energy states. Finally, upon increasing the size from MODEL-3 to MODEL-4, both CASSCF and NEVPT2 show signs of convergence: MODEL-4 reproduces the MODEL-3 energy levels within 0.1 eV for the interesting electronic states (see Table S3).
In line with this,  we also found that the defect orbitals have marginal weight at the terminating hydrogens for MODEL-3 and 4 contrary to the smaller cluster models (see section S2.3 in SI). Based on the spotted tendencies, we conclude that MODEL-4 is already large enough to accommodate the defect center properly and the inclusion of an additional shell (i.e. MODEL-5) is not expected to produce energy contributions over 0.1 eV.

As for further parameters of interest, the largest difference between MODEL-3 and MODEL-4 results - which defines the error margin deriving from finite model and basis size - was found to be 0.015 \AA{} for bond distances in equilibrium geometries (see SI, Table S4 and S5), 7 meV for Jahn-Teller stabilization energies and barriers (Table S9), and 0.05 GHz for zero-field spitting parameters (D tensor of ground triplet state, $D_{es}$ and $\Delta$ tensors of the first excited state; see Table S10 in SI). The only quantity where significant uncertainty was observed was the spin-orbit coupling induced splitting of triplet $E$ states ($\lambda_z$) with 1.3 GHz of alteration between MODEL-3 and MODEL-4. This exception, however, primarily derives from the exponential dependence of Ham reduction factors on computable properties (see the related subsections for details). 

Altogether, the convergence analysis with respect to model size and basis set size indicates that the CASSCF-NEVPT2 results presented in this manuscript (obtained using MODEL-4 and cc-pVDZ basis) are expected to contain negligible errors that stem from finite-size effects.

\bigbreak
\noindent {\bf Defect orbitals}

Since the results have been demonstrated to be convergent at the cluster size of MODEL-4, the main text discusses on the data obtained with MODEL-4 from this point on. We begin our investigation by examining the Hartree–Fock orbitals of the defective nanodiamond, computed for the $^3\!A_2$ ground state. We identified the four defect orbitals of $sp^3$ hybrid character  of the nanodiamond, see illustratation in Fig.~\ref{fig:Configurations} (top). These orbitals correspond to the dangling bonds of the three carbons and the nitrogen adjacent to the vacancy. Among these, the $a_1$ is a bonding orbital shared between the three carbon and nitrogen atoms, while $a_1^*$ is its anti-bonding pair. The degenerate orbitals $e_x$ and $e_y$, which are responsible for the spin-density  distribution, are localised exclusively on the carbon atoms. By studying the Löwdin orbital composition of these defect orbitals, we confirmed numerically that all of them are primarily concentrated in the inner shells of the models. For more details consult section S2.3 in SI. 

The four defect orbitals described above are occupied by six valence electrons, the distribution of which is expected to characterize the low-energy spectrum\cite{Maze2011}. The electron configurations of the conceivable 6 triplet and 10 singlet states of the defect states, determined through group theoretical considerations\cite{Maze2011}, are presented on the bottom part of Fig.~\ref{fig:Configurations}. While the spin-polarization loop of NV$^-$ is known\cite{Thiering_2018} to proceed through the six lowest-lying electronic states of $(1){^3}\!A_2$, $(1){^3}\!E_x$, $(1){^3}\!E_y$, $(1){^1}\!E_x$, $(1){^1}\!E_y$ and $(1){^1}\!A_1$, higher-lying states within the band gap are also worth studying. Namely, recent transient absorption spectroscopy experiments\cite{Luu2024} revealed hitherto uncharacterized optical transitions, which can only be explained by localized states lying above $(1)^3E$ and $(1)^1A_1$ in the triplet and singlet sector, respectively.

We emphasize here that the depicted character of each state can be significantly altered by state mixing effects, which cannot be quantitatively predicted by group theory. However, our many-body WFT-based protocol takes state mixing into account by construction. 

As a conclusion of the  above findings, we form our minimalist CASSCF active space of the four defect orbitals\cite{Maze2011,Doherty2011}.

\begin{table*}[t]
\setlength\extrarowheight{3pt}
\centering
\resizebox{\textwidth}{!}{%
\begin{threeparttable}
\caption{Comparison of CASSCF-NEVPT2 vertical energy levels (i.e. energy differences excluding geometry relaxation effects) to previously reported theoretical values. In the table, the many-body $(1)^1\bar{E}$ and $(1)^1\bar{A}_1$ energies are referenced to the $^3\bar{A}_2$ ground state energy,  $\Delta_S$ and $\Delta_T$ denote the singlet  and triplet transition energies, i.e., $(1)^1\bar{E} \rightarrow (1)^1\bar{A}_1$  and $^3\bar{A}_2 \rightarrow ^3\bar{E}$. In the calculations referenced for this work, we applied the CASSCF(6e,4o)-NEVPT2 approach. SA($m+n$) scheme denotes results where the lowest $m$ triplet and $n$ singlet states were considered in the state-averaging, while SA($X$,$Y$) scheme denotes state-averaged computations involving only $X$ and $Y$ states.}
\begin{tabular}{|c|c|c|c|c|c|c|c|c|}
\hline
&&&& \multicolumn{4}{|c|}{Vertical energy [eV]} &\\
\hline
Methodology & Level of theory & Model & \# carbons &  $(1)^1\bar{E} $ & $(1)^1\bar{A}_1$ & $\Delta_S$ & $\Delta_T$ &  Reference \\
\hline
\hline
Experiment & - & - & - & - & & 1.26 & 2.18 & \citenum{Kehayias_2013,Davies:PRSLA1976}\\
\hline
WFT & SA(5+8)-scheme & Cluster & 145 & 0.56 & 1.60 & 1.04 & 2.18 & This work \\
& SA(3+3)-scheme & Cluster & 145  & 0.62 & 1.77 & 1.15 & 2.35 & This work \\
& SA($(1)^3A_2, (1)^3E$)-scheme & Cluster & 145 & - & - & - & 2.31 & This work \\
& SA($(1)^1E, (1)^1A_1$)-scheme &  Cluster & 145  & - & - & 1.13 & - & This work \\
& SA-CASSCF(6e,6o) &  Cluster & 85 & 0.25 & 1.60 & 1.35 & 2.14 & \citenum{Bhandari2021} \\
& SA-CASSCF(6e,6o) &  Cluster & 33 & 0.66 & 1.96 & 1.30 & 2.30 & \citenum{li2024excitedstatedynamicsopticallydetected}\\
& SA-CASSCF(6e,6o)-CASPT2 &  Cluster & 33 & 0.55 & 1.57 & 1.02 & 2.22 & \citenum{li2024excitedstatedynamicsopticallydetected}\\
& MCCI &  Cluster & 42 & 0.63 & 2.06 & 1.43 & 1.93 & \citenum{Delaney2010}\\
& NEVPT2-DMET &  Supercell & 212 & 0.50& 1.52& 1.02 & 2.31 & \citenum{Haldar_2023} \\
\hline
WFT in DFT & FCIQMC-in-DFT &  Cluster & 42 & 0.58 & - & - & 1.98 & \citenum{Chen_2023} \\
& CI-cRPA & Supercell & 510 & 0.49& 1.41& 0.92 & 2.02 & \citenum{Bockstedte2018} \\
& beyond-RPA & Supercell & 214 & 0.56& 1.76& 1.20 & 2.00 & \citenum{Ma2020} \\
& QDET by Green's function & Supercell & 510 & 0.46& 1.27& 0.81 & 2.15 & \citenum{Sheng_2022} \\
\hline
MBPT on DFT & GW+BSE & Supercell & 254 & 0.40 & 0.99 & 0.59 & 2.32 & \citenum{Ma2010} \\
 & GW + extended Hubbard & Supercell  & & 0.5 & 1.5 & 1.0 & 2.1 & \citenum{Choi2012} \\
  & SF-BSE & Supercell  & & 0.44 & 1.05 & 0.61 & 2.07 & \citenum{Barker2022} \\
\hline
DFT & BP86 &  Cluster & 284 & 0.48 & 2.03 & 1.55 & 1.90 & \citenum{Delaney2010} \\
& PZ81 & Supercell & 510 & - & - & - & 1.91 & \citenum{Gali2008} \\
& HSE06 & Supercell & 510 & - & - & - & 2.21 & \citenum{Gali:PRL2009} \\
& DO-MOM($r^2$SCAN) & Supercell & 510 & 0.62 & 1.80 & 1.18 & 2.06 & \citenum{Ivanov_2023} \\
& TDDFT(PBE) & Supercell & 214 &0.52 & 1.38 & 0.86 & 2.08 & \citenum{Jin2022} \\

\hline
\end{tabular}
\label{table:Theory_comparison}  
\end{threeparttable}}
\end{table*}

\bigbreak
\noindent {\bf Vertical electronic spectrum}
\begin{figure*}[t!]
\begin{center}
\includegraphics[width=\textwidth]{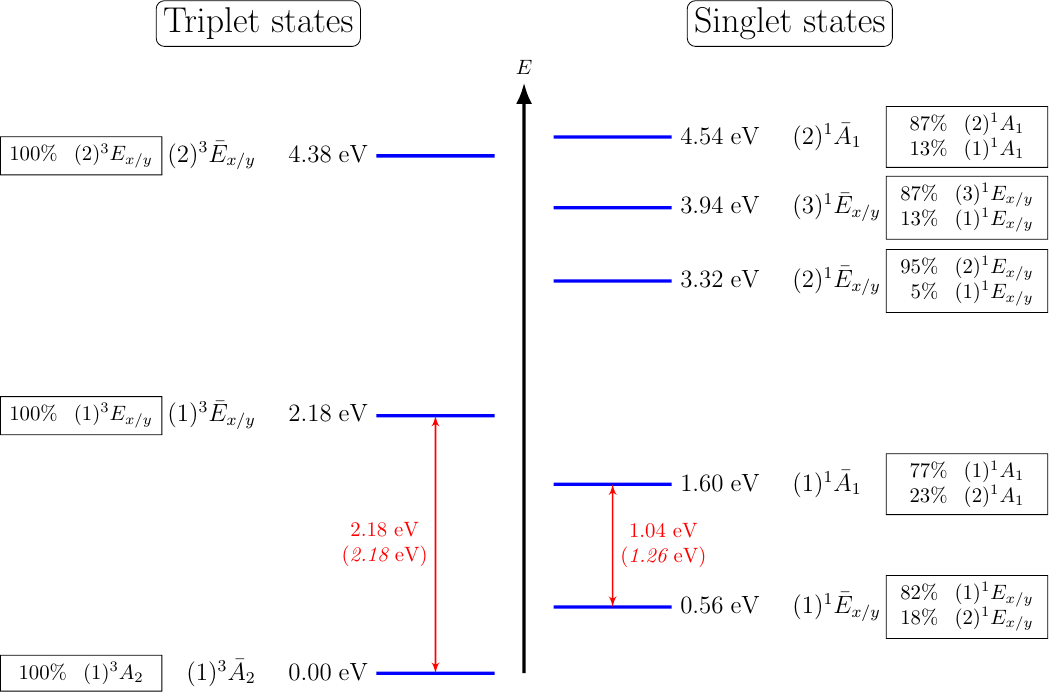}
\caption{Vertical electronic spectrum of the NV- center, as calculated at SA(5+8)-CASSCF(6e,4o)-NEVPT2/cc-pVDZ level of theory, at the geometry of the ground ($^3A_2$) state of MODEL-4. Red arrows indicate experimentally characterized vertical energies, while plain and parenthesized numbers stand for calculated and measured vertical energy difference, respectively. (We note that the effect of number of states in the state-averaging slightly affects the calculated results, see Table \ref{table:Theory_comparison}).Black frames show the composition of electronic states, expressed in the basis of pure (group-theory) configurations presented in Fig. \ref{fig:Configurations} Determinants with weights over 1\% in the CASSCF wavefunction are provided, and the most simple wavefunction representations are shown, taking into account that $a_1$ and $a_1^*$ molecular orbitals can be arbitrarily combined at $C_{3v}$ symmetry. }
\label{fig:Vertical}  
\end{center}
\end{figure*}

In accordance with the presented defect orbital picture, we proceeded with an initial analysis of electronic states at ground-state geometry, using CASSCF(6e,4o) calculations state-averaged for all conceivable (6+10) roots. The vertical energies obtained in this manner for MODELs 1-4 can be found in SI, Table S2. The analysis of these data revealed that the highest-energy states, specifically $(1)^3A_1, (3)^1A_1$ and $(4)^1A_1$ are intensively mixed with the conduction band and hence cannot be considered as band gap states. This can be seen from (i) their outstandingly high CASSCF-NEVPT2 vertical energy level ($>$ 5 eV), (ii) the negative denominators appearing during NEVPT2 calculation, and (iii) the fact that the CASSCF-NEVPT2 energy levels of the 16 electronic states show slow convergence with respect to model size, {and change significantly even between MODEL-3 and MODEL-4}. The latter phenomenon is the result of the deterioration of state-averaged CASSCF orbitals due to the presence of the three aforementioned unphysical states. Thus, we decided to decrease the number of states and repeated the vertical energy calculations with 5 triplet and 8 singlet roots, which gave well-converged results (see SI, Table S3).

We summarize the composition of the 5 triplet and 8 singlet CASSCF eigenstates, indicated by an overline in the following,  in the basis of the electron configurations derived from group theory, see black frames in Fig.~\ref{fig:Vertical}. Conspicuously, the triplet states are clearly of single-reference, meaning they can be effectively described by a single electron configuration. On the other hand, the singlet states arise as a mixture of multiple group-theory configurations. The presence of multiple dominant Slater determinants cannot be treated by conventional DFT methods, as they can only be generated from each other by electron excitation(s) between orbitals of different irreducible representations.

For instance, the $(1)^1\bar{A}_1$ CASSCF eigenstate  contains not only the dominant $(1)^1A_1$ configurations proposed by group theory (76\%) but also the $(2)^1A_1$ configurations with an empty $a_1^*$ orbital (24\%). Furthermore, the $(1)^1\bar{E}$ CASSCF eigenstates admix $a_1^* \rightarrow e$  excitations ($(2)^1E$ character, 19\%) to the group-theoretic pure $(1)^1E$ states (81\%). The higher-energy states of $(2)^1\bar{E}, (3)^1\bar{E}$ and $(2)^1\bar{A}_1$ also carry the character of the lowest-lying singlet state, but their multireference character is slightly less apparent (8-13\%). 

Investigating the determinant composition is also useful for identifying the potential mixing of electronic states by phononic coupling (pseudo Jahn-Teller effect). For example, it can be immediately recognized from Fig. \ref{fig:Configurations} that $(1)^1A_1$ and $(1)^1E_y$ (and, consequently, $(1)^1\bar{A}_1$ and $(1)^1\bar{E}_y$) share two leading Slater determinants. This implies that considerable state mixing is expected upon breaking the $C_{3v}$ symmetry, which manifests - among others - in a rapid internal conversion (IC) rate. In the triplet sector, on the contrary, the dominant $(1)^3\bar{A}_2$ configuration is orthogonal to the leading determinants of all $^3\bar{E}$ states because the group theoretical states do not mix according to CASSCF. This suggests that state mixing is negligible even considering vibrations and $(1)^3\bar{E}$ is stable to internal conversion. Altogether, the determinant picture is fully consistent with the experimental observation\cite{Rogers-2008} that the photoluminescence between singlets is $\sim 10^4$ times weaker than that between triplets, since the latter is not suppressed by IC.  

Now, we turn the focus to the discussion of the vertical excitation energies in Fig.~\ref{fig:Vertical}. Analyzing the numerical data in Table S3 of SI, we observed that the application of the NEVPT2 correction on top of the CASSCF theory had a large impact by reducing the raw CASSCF gaps by approximately 50\% in average. This decrease corresponds to 0.4-6.2 eV alteration in absolute value (see SI, section S3 for details). Overall, the resulting CASSCF-NEVPT2 energy spectrum for MODEL-4 is in fair qualitative agreement with the experimentally observed optical transitions, in spite of the compromised orbital optimization among many states and the lack of geometry relaxation effects. Specifically, we predict 2.18 and 1.04 eV for the lowest vertical triplet and singlet optical gaps (corresponding experimental zero-phonon lines: 1.95 for $(1)^3\bar{E}\rightarrow(1)^3\bar{A}_2$ and 1.12 eV $(1)^1\bar{A}_1\rightarrow(1)^1\bar{E}$). In addition, the position of $(2)^3\bar{E}$ (2.20 eV above $(2)^3\bar{E}$) in the triplet sector, and that of $(2)^1\bar{E}$ (1.82 eV above $(1)^1\bar{A}_1$), $(3)^1\bar{E}$ (2.34 eV above $(1)^1\bar{A}_1$) and $(2)^1\bar{A}_1$ (2.94 eV above $(1)^1\bar{A}_1$) in the singlet sector can approximately correspond to the newly observed transitions of NV$^-$ discovered by transient absorption spectroscopy\cite{Luu2024}. 
The latter experiment, for the first time, reported one (2.38 eV) and four (1.75, 2.78, 2.79 and 2.85 eV) absorption peaks excited from $(1)^3\bar{E}$ and $(1)^1\bar{A}_1$ states, respectively.

As structural relaxation was often neglected in previous theoretical studies on the energy spectrum of NV$^-$, it is interesting to compare our vertical energy values of the lowest-energy states ($(1)^3\bar{A}_2: 0.00$ eV; $(1)^1\bar{E}: 0.56$ eV; $(1)^1\bar{A}_1: 1.60$ eV; $(1)^3\bar{E}: 2.18$ eV) to earlier works. The comparison is summarized in Tab.~\ref{table:Theory_comparison}. As a preliminary remark, we note that "experimental" vertical energies can also be estimated by locating the maximum-intensity peak of the phonon sideband spectra, though these values are somewhat ambiguous. For the first triplet and singlet vertical gap, 2.18 eV and 1.26 eV, respectively, were measured in this manner. While the former value exactly matches our results (2.18 eV, 1.04 eV), the latter is slightly off by $\approx0.2$ eV. (See also Fig. \ref{fig:Vertical}, bottom.) For completeness, we add that changing the weighing in orbital optimization has a slight but notable effect on the calculated energies - for example, if we restrict the state averaging for three triplet and three singlet states (CASSCF(3+3)), the NEVPT2 singlet gap (1.15 eV) gets closer to the experimental reference value, but the triplet gap is also somewhat shifted upwards, to 2.35 eV. Similar results (1.13 eV and 2.31 eV, respectively) were obtained by optimizing the orbitals specifically for the two states between which the gap is to be calculated  (see also Tab.~\ref{table:Theory_comparison}, upper rows). The latter way of computation gives - in theory - the most reliable vertical gaps.

Regarding previous theoretical studies, the gap between $(1)^3\bar{A}_2$ and $(1)^3\bar{E}$ was reasonably calculated even by pure DFT, as only single-reference electronic states are involved. For example, energy differences of 1.90-1.91 eV were calculated by BP86 and PZ81 functionals, applied to a cluster model and a supercell, respectively\cite{Gali2008,Delaney2010} while supercell HSE06 calculation resulted in 2.21 eV\cite{Gali:PRL2009}. Post-DFT methods, e.g., CI-cRPA: 2.02 eV\cite{Bockstedte2018}, beyond-RPA: 2.00 eV\cite{Ma2020}, GW: 2.1-2.3 eV\cite{Ma2010,Choi2012}, approach the experimental energy difference  better, albeit the deviation from pure DFT is rather marginal. Among wavefunction-based results, the CASSCF results with an extended active space (6 electrons on 6 orbitals, i.e. two local virtual orbitals added to (6e,4o)) are particularly noteworthy: $(1)^3\bar{E}$ energy levels of 2.1-2.3 eV were obtained without dynamic correlation treatment, depending on the way of state-averaging and the cluster size (34-86 heavy atoms)\cite{Bhandari2021,li2024excitedstatedynamicsopticallydetected}. CASPT2 on top of the latter CASSCF(6e,6o) wavefunctions resulted in a gap of 2.22 eV\cite{Bhandari2021,li2024excitedstatedynamicsopticallydetected}. Recently developed embedding methods, such as quantum defect embedding theory (QDET; $\Delta E = 2.15$ eV\cite{Sheng_2022}) and density matrix embedding theory using NEVPT2 (NEVPT2-DMET; $\Delta E = 2.31$ eV\cite{Haldar_2023}) also provided similar values. Altogether, our CASSCF-NEVPT2 results of 2.18-2.35 eV (depending on the weighing) are consistent with multiple versatile methodologies.

In contrast to the triplet gap, the energy levels of the singlet states show remarkable method dependence. For example, QDET\cite{Sheng_2022} and GW+BSE\cite{Ma2010} provide outstandingly low energies for $(1)^1\bar{A}_1$ (resulting in underestimated singlet gaps of 0.81 and 0.59 eV, respectively). On the other hand, some other approaches - such as DFT\cite{Gali2008,Delaney2010} and small-cluster Monte Carlo configuration interaction (MCCI, FCIQMC)\cite{Delaney2010,Chen2023} - tend to place $(1)^1\bar{A}_1$ (with $E\approx 2.0$ eV) over $(1)^3E$ (with $E\approx 1.9$ eV), which contradicts the experimentally observed facile triplet-to-singlet intersystem crossing. Importantly, our CASSCF-NEVPT2 methodology does not produce any of the latter erroneous results and gives a $(1)^1\bar{A}_1$ energy level of 1.60 eV, in perfect agreement with the closely related methods (CASPT2: 1.57 eV\cite{li2024excitedstatedynamicsopticallydetected}, NEVPT2-DMET: 1.52 eV\cite{Haldar_2023}). While certain methods reproduce the experimental gap of 1.26 eV more closely (such as beyond-RPA\cite{Ma2020} with 1.20 eV and quite surprisingly SA-CASSCF(6e,6o)\cite{li2024excitedstatedynamicsopticallydetected} with 1.30 eV), the CASSCF-NEVPT2 energy difference between singlets (1.04-1.15 eV) still has a tolerable error.

\bigbreak
\noindent {\bf Relaxed electronic spectrum}
\begin{figure*}[t]
\begin{center}
\includegraphics[width=\textwidth]{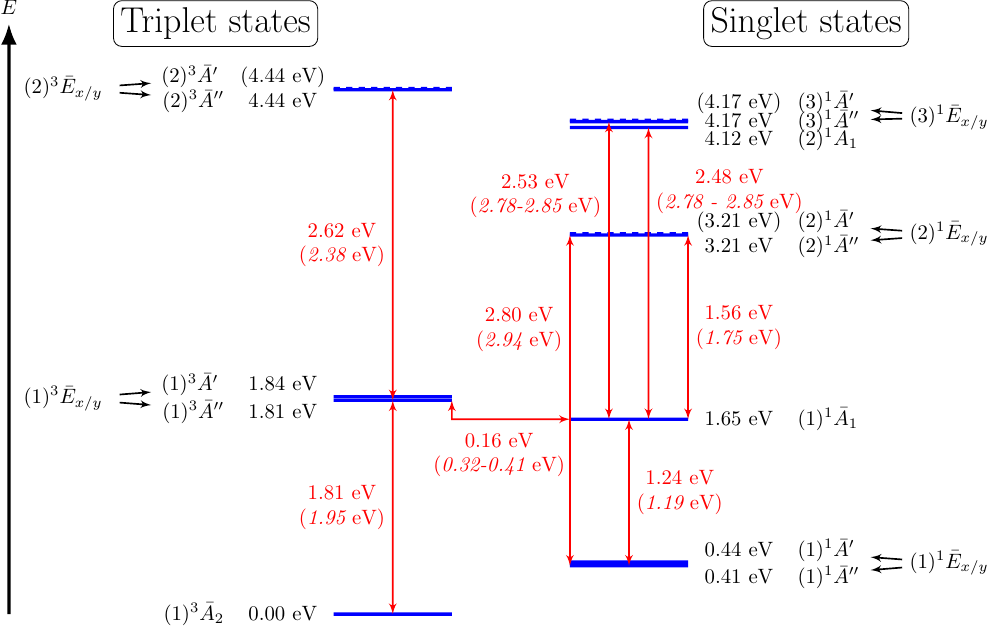}
\caption{Relative energy levels of the electronic states of the NV$^-$ center (SA-CASSCF(6e,4o)-NEVPT2/cc-pVDZ), including geometry relaxation effects (i.e. geometry optimization at SS-CASSCF level for each state). The 5 lowest triplet and 8 lowest singlet CASSCF eigenstates were taken into account. Red arrows indicate experimentally characterized optical transitions, where the computational data can be directly compared to measurements. Plain and parenthesized numbers indicate theoretical and experimental energy differences, respectively.}
\label{fig:Relaxed}  
\end{center}
\end{figure*}

To determine adiabatic energy differences between the states that correspond to the experimental zero-phonon lines, we optimized the geometry of each state using the state-specific CASSCF(6e,4o)/cc-pVDZ method. Then, accurate electronic energy differences were computed for the relaxed geometries at SA-CASSCF(6e,4o)-NEVPT2/cc-pVDZ level of theory. To obtain the highest possible accuracy, the state-averaging in the latter step only involved the two states of interest (rather than all 5 + 8 electronic states) for each pair of states and the full energy spectrum was assembled by summing the individual energy differences in a post-process manner.

The excited-state energy levels including geometry relaxation effects are summarized in Fig.~\ref{fig:Relaxed}.  The equilibrium geometry for each electronic state, characterized by the C-C distances adjacent to the vacancy, is summarized in the SI, Table S5. Furthermore, a detailed description of the calculation of relative energy levels, as well as the discussion of model size dependence is also provided in section S3.3 of SI.

Concerning the state-specific equilibrium geometries, we find that states of $A_1$ and $A_2$ irreducible representations conserve the $C_{3v}$ symmetry even after the geometry optimization procedure. Nevertheless, the C-C distances adjacent to the vacancy do depend on the electronic structure; for example, in the excited singlet ($(1)^1\bar{A}_1$) state of the NV$^-$ polarization circle, the carbon atoms are $\approx0.1~\mathrm{\AA}$ further from each other than in the ground ($(1)^3\bar{A}_2$) state.  

As for the $\bar{E}$ states, we clearly observe the expected Jahn-Teller distortion, which lowers the symmetry to $C_{1h}$. This splits the double degeneracies observed in the vertical energy spectrum and forms $\bar{A}'$ and $\bar{A}''$ states of slightly different geometries and - consequently - energy levels (Fig.~\ref{fig:Relaxed}, black arrows). In the former case, one inner C-C distance elongates relative to the other two, but the system remains close to $C_{3v}$ symmetry - only 0.02-0.03 \AA{} deviations were observed. In the latter case, elongation occurs, and the two longer C-C distances exceed the short C-C side by 0.09-0.14 \AA.  Importantly, the energy difference between $\bar{A}^{\prime}$ and $\bar{A}^{\prime\prime}$ states falls in the range of tens of meVs, which foreshadows that dynamic Jahn-Teller effect might be present. A more sophisticated discussion of JT energy levels and dynamic JT behavior can be found in the next section. (For high-energy states, only $\bar{A}''$ Jahn-Teller minima could be optimized due to technical issues, i.e. uncontrollable root flipping between neighboring $\bar{A}'$ states. Nevertheless,
assuming that the energy of JT stabilization is similar for the two irreducible representations, it is sufficient to characterize the energy level of these states based on merely the $A''$ branch.)   

Next, we evaluate the relative energy levels of the electronic states, which are  directly comparable to experimental zero-phonon lines (ZPL). In the triplet spin sector (Fig.~\ref{fig:Relaxed}, left) the well-known characteristic ZPL of NV$^-$ (1.95 eV energy change corresponding to 637 nm emission), is reproduced as 1.81 eV energy difference between $(1)^3\bar{E}$ and $(1)^3\bar{A}_2$. Moreover, the equilibrium geometry of the second excited triplet state ($(2)^3\bar{E}_x\rightarrow^3\bar{A}''$) lies at 4.44 eV above the ground state, that is, 2.62 eV above the corresponding first excited state of $(1)^3\bar{E}_x\rightarrow^3\bar{A}''$. The difference reasonably agrees with the experimentally observed 2.38 eV signal in transient absorption spectroscopy\cite{Luu2024} deriving from the excitation of $^3\bar{E}$ to a state with unidentified symmetry. We note that the enhanced error of the CASSCF-NEVPT2 excitation energy between $(1)^3\bar{E}$ and $(2)^3\bar{E}$ (2.62 - 2.38 = 0.24 eV), relative to the error for the lower optical transition (1.95 -1.81 = 0.14 eV) is not surprising, as CASSCF excited states are computed as orthogonal vectors to all lower states, the numerical uncertainties of which accumulate to a higher and higher extent with the increasing number of the required roots. Still, even the deviation of 0.24 eV can be considered as a close reproduction of the measured data, taking into account that the energy gap between the electronic states is in the range of 2 eV and that dynamic correlation effects are computed at perturbative level.

As for the singlet states (Fig.~\ref{fig:Relaxed}, right), the optical transition belonging to the infrared emission of NV$^-$ (1042 nm, 1.19 eV) is accurately modeled by CASSCF-NEVPT2 - 1.24 eV energy gap is obtained after relaxation. Above the familiar excited singlet state of $(1)^1\bar{A}_1$, the lowest accessible excited state was calculated to be $(2)^1\bar{E}$. At CASSCF-NEVPT2 level, this state lies 1.56 eV higher than $(1)^1\bar{A}_1$ and 2.80 eV higher than the lowest-energy singlet state ($(1)^1\bar{E}$), which is in good agreement with the experimentally observed signals of 1.75 and 2.94 eV\cite{Luu2024}, respectively. Among the singlet states localized to the NV$^-$ center, $(3)^1\bar{E}$ and $(2)^1\bar{A}_1$ remain, which calculated energy levels of 2.48-2.53 eV compared to $(1)^1\bar{A}_1$. These numbers can be corresponded to the transient absorption signals of 2.78, 2.79 and 2.85 eV. The deviation of $\approx 0.3$ eV is tolerable for such high-energy states, though it remains an open question why three signals are seen in the aforementioned range. (Although $(3)^1\bar{E}_x$, $(3)^1\bar{E}_y$ and $(2)^1\bar{A}_1$ represent three possible excited states, the two quasi-degenerate $E$ states typically produce one sharp peak in absorption spectra.)    

Altogether, our CASSCF-NEVPT2 methodology not only reproduced the relative energy levels of the familiar states of the polarization cycle, but it also predicted high-energy band-gap states with reasonable agreement with recent transient absorption spectroscopy data. The average absolute and relative error of CASSCF-NEVPT2, compared to zero-phonon line energies of experimentally observable optical transitions, was 0.20 eV and 8.6\%, respectively. Considering only the low-energy states up to $(1)^3\bar{E}$ and $(1)^1\bar{A}_1$, the agreement is even better (0.09 eV, 5.5\%).

\bigbreak
\noindent {\bf Pressure dependence of ZPL}

\begin{figure}[!t]
\begin{center}
\includegraphics[width=0.47\textwidth]{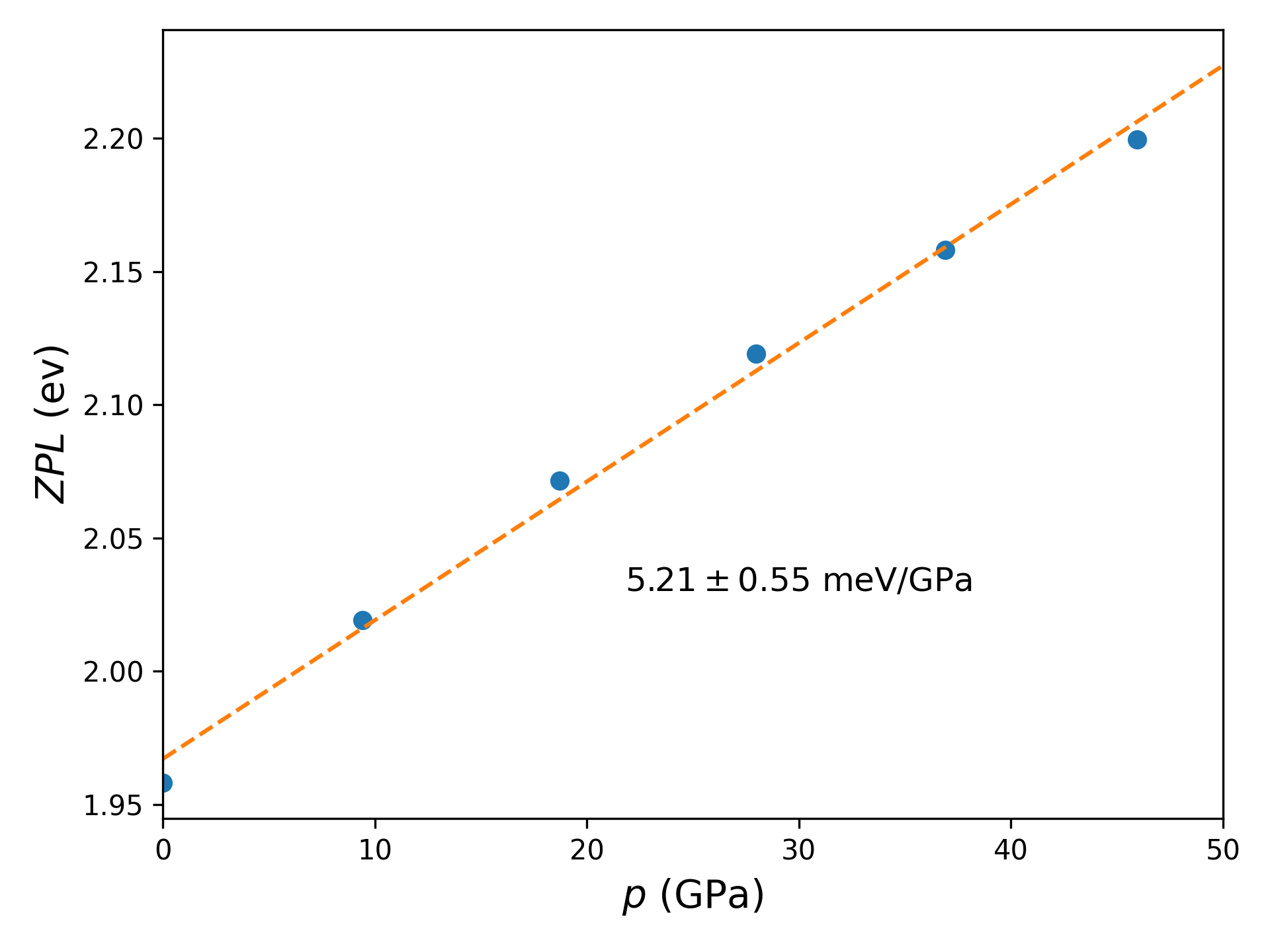}
\caption{Dependence of the zero-phonon line of $(1)^3\bar{E}\rightarrow(1)^3\bar{A}_2$ photoluminescence on the hydrostatic pressure. The ZPL was calculated as energy difference between the CASSCF eigenstates of $(1)^3\bar{A}_2$ and $(1)^3\bar{E}$, including NEVPT2 correction and constrained geometry relaxation effects. }
\label{fig:pressure}  
\end{center}
\end{figure}

The NV center has a potential in quantum metrology applications due to its sensitivity to external fields and surrounding conditions, e.g. temperature and pressure.
In particular, the ZPL between the lowest triplets was shown to vary linearly with respect to hydrostatic pressure\cite{Kobayashi_1993,Doherty_2014}.
Here, we aimed to reproduce the experimental observations in order to demonstrate the general applicability of our proposed modeling scheme.
In order to simulate the effect of the  hydrostatic stress in our cluster model, we repeated the geometry construction procedure  discussed above (see  Fig.~\ref{fig:models}) but applying bond length parameters tuned according to the numerical pressure-density\cite{Guler_2015} and the trivial density-bond length relationships. 
In practice, we studied MODEL-3 whose geometry  was initialized by rescaling the atomic coordinates corresponding to  pressure and it was optimized keeping the two outermost atomic shells frozen while maintaining  $C_{3v}$ symmetry.

The pressure dependent ZPL data, visualized in Fig.~\ref{fig:pressure}, were calculated as CASSCF-NEVPT2 energy differences between $(1)^3\bar{E}$ and $(1)^3\bar{A}_2$ in the 0-50 GPa pressure range.  By applying linear fitting to the numerical data, we obtain a $5.21\pm 0.55$ meV/GPa gradient which is in good agreement with experimental findings\cite{Kobayashi_1993,Doherty_2014} of 5.5-5.75 meV/GPa. This result gives further evidence for the validity of the cluster based modeling with constrained geometry optimization.

\bigbreak
\noindent {\bf Jahn-Teller behavior of $E$ states}
\begin{figure}[!t]
\begin{center}
\includegraphics[width=0.5\textwidth]{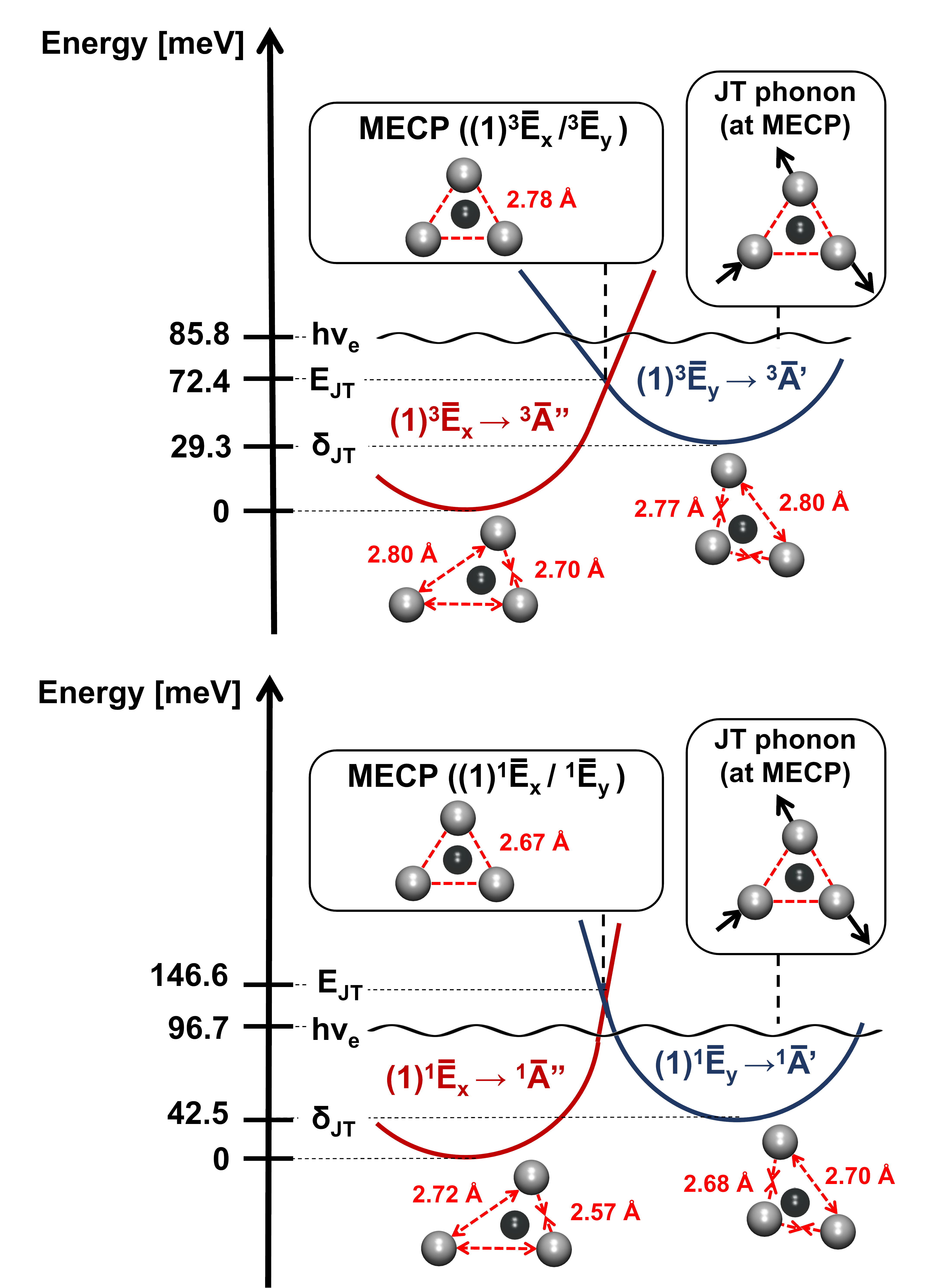}
\caption{Schematic representation of potential energy hypersurfaces in the vicinity of the minimum-energy crossing points (MECPs) between $(1)^3\bar{E}$ states (upper part) and $(1)^1\bar{E}$ states (bottom part). The relative energy levels of $\bar{A}^{\prime}$, $\bar{A}^{\prime \prime}$ and MECP structures were calculated at SS-CASSCF(6e,4o)-NEVPT2 level. Jahn-Teller parameters ($E_{\rm JT}, \delta_{\rm JT}$) were obtained from energy differences, as shown. Parameter $h\nu_e$ was calculated as an effective phonon energy between $\bar{A}'$ and $\bar{A}''$ minima, using 50-50\% weighing for the two degenerate $\bar{E}$ states. In the structure plots, the carbon atoms neighboring the vacancy are presented, N-V axis is out of plane.}
\label{fig:JT}  
\end{center}
\end{figure}

When studying the Jahn-Teller distorted $^3\bar{E}$ and $^1\bar{E}$ states, it is of fundamental interest whether the system is trapped in a single potential energy valley corresponding to an energetically highly favorable $\bar{A}^{\prime \prime}$ or $\bar{A}^{\prime}$ structure (static Jahn-Teller effect), or continuously oscillates among three spatially degenerate minima (dynamic Jahn-Teller effect). Namely, in the latter case, the system appears to be of high ($C_{3v}$) symmetry in experiments, even if the energetically most favorable geometry is distorted. In this section, the JT behavior of the low-energy $(1)^3\bar{E}$ and $(1)^1\bar{E}$ states will be discussed. 

Static and dynamic Jahn-Teller systems can be distinguished by comparing the Jahn-Teller barrier ($\delta_{\rm JT}$) to the zero-point energy level of the two distortion-driving $e$ vibrational modes ($h\nu_e$). Here, $\delta_{\rm JT}$ refers to the energy difference between the two $C_{1h}$ symmetrical configurations, i.e. $A^{\prime \prime}$ and $A^{\prime}$ (one of them appears as energy minimum, while the other acts as a transition state in Jahn-Teller oscillation). The criterion for a static JT effect is $\delta_{\rm JT} > h\nu_e$; in the opposite case, the energy of the vibration is sufficiently large to form a dynamic JT system\cite{Bersuker_2006}. As visualized in Fig.~\ref{fig:JT}, $\delta_{\rm JT}$ can be calculated as an energy difference between $\bar{A}^{\prime \prime}$ and $\bar{A}^{\prime}$ geometries.

For further investigations, we searched for the minimum-energy crossing point (MECP) between $\bar{E}$ states at CASSCF level, by setting the weight of both $\bar{E}_x$ and $\bar{E}_y$ to 50\% in the CASSCF optimization run. The resulting MECP (geometry for $(1)^3\bar{E}$ and $(1)^1\bar{E}$ is depicted in Fig.~\ref{fig:JT}, see black frames) is the most stable geometry where the $\bar{E}_x$ and $\bar{E}_y$ are degenerate, which is only possible at $C_{3v}$ symmetry. As indicated in Fig.~\ref{fig:JT}, the energy difference between MECP and the bottom of the lowest-energy valley corresponds to the Jahn-Teller stabilization energy ($E_{\rm JT}$), the role of which in the modeling is discussed below.  

As MECP is present in both $\bar{A}'$ and $\bar{A}''$ potential energy surfaces (as shown in Fig.~\ref{fig:JT}), it is possible to calculate the relative energy levels of the characteristic geometries ($\bar{A}'$ minimum, $\bar{A}''$ minimum and MECP) using two state-specific CASSCF calculations. We find that the MECP $\rightarrow$ $\bar{A}''$ energy gain (calculated with $\bar{A}''$-specific CASSCF, corresponding to the red PES) is consistently higher than that for MECP $\rightarrow$ $\bar{A}'$ (calculated with $A'$-specific CASSCF, corresponding to the blue PES), which implies that $\bar{A}''$ is always the global JT minimum. Even though this result is somewhat in line with chemical intuition (i.e. the occupation of covalent C-C bond forming $e_x$ orbitals is higher in $\bar{A}''$ states),  it contradicts previous DFT results\cite{thiering_ab_2017}, according to which $\bar{A}'$ is slightly more favorable for $(1)^3E$. 
The apparent ambiguity might just highlight the limitations to energy resolution of the methods with different hierarchy of approximations.

As for the numerical data, we obtain $\delta_{\rm JT}$ = 29 and 43 meV for $(1)^3\bar{E}$ and $(1)^1\bar{E}$, respectively (see detailed discussion and convergence behavior with respect to model size in SI, section S4). To determine the nature of the JT effect, these values should be compared to the energy of the effective $e$ phonon\cite{Alkauskas2014} that connects $\bar{A}'$ and $\bar{A}''$ minima. Using a linear PES scan at CASSCF level with 50-50\% weighing for $\bar{E}_x$ and $\bar{E}_y$ (which mimics the vibration behavior before the manifestation of the JT effect; see SI, section S4.1 for details), $h\nu_e$ = 86 meV and 97 meV was found for $(1)^3\bar{E}$ and $(1)^1\bar{E}$, respectively. Therefore, both $(1)^3\bar{E}$ and $(1)^1\bar{E}$ will be investigated as dynamic JT systems in the following.

In dynamic JT, the strength of the vibronic coupling, i.e., the proportion of the Jahn-Teller stabilization energy ($E_{\rm JT}$) and the Jahn-Teller phonon energy ($h\nu_e$) characterizes the behavior of the system; it determines how the observed properties, such as the fine structure (\emph{vide infra}) is altered compared to the Born-Oppenheimer picture.

$E_{\rm JT}$ refers to the energy which is released upon symmetry breaking ($C_{3v} \rightarrow C_{1h}$). It equals to the energy level of MECP relative to the JT minimum. At SS-CASSCF-NEVPT2 level, we found $E_{\rm JT}$ = 72 and 147 meV for $(1)^3\bar{E}$ and $(1)^1\bar{E}$, respectively. Putting these data together, both systems turn out to be strongly coupled, as $h\nu_e$ and $E_{\rm JT}$ are commensurable. 

\bigbreak
\noindent {\bf Fine structure of triplet states}

\begin{figure*}[!t]
\begin{center}
\includegraphics[width=\textwidth]{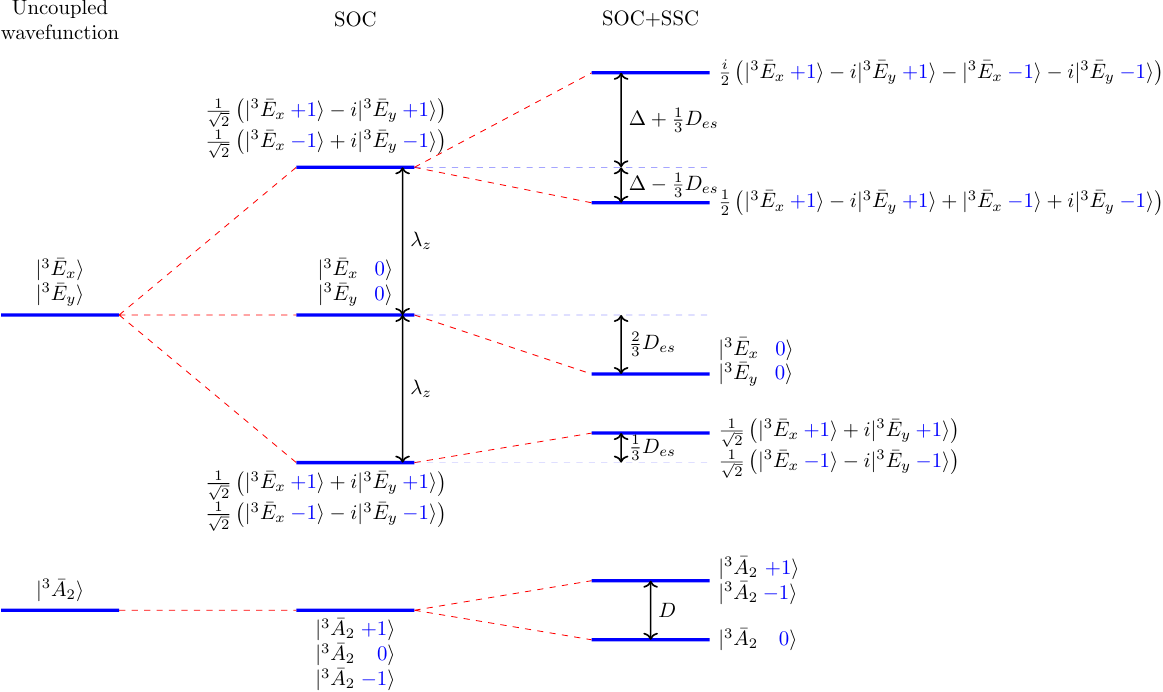}
\end{center}
\caption{Visualization of the splitting of $(1)^3\bar{A}_2$ and $(1)^3\bar{E}$ states due to spin-orbit and spin-spin  coupling. At each SOC (and SSC) coupled state, we provide the wavefunction on the basis of spin sublevels (blue numbers stand for $m_S$ quantum number). After the QDPT treatment of the CASSCF wavefunctions, which introduces SOC and SSC couplings in a post-process manner, we extracted the indicated parameters ($D$, $\lambda_z$, $D_{es}$, $\Delta$) from the energy differences between QDPT eigenstates.}
\label{fig:ZFS}
\end{figure*}
\begin{table}[b]
\setlength\extrarowheight{4pt}
\centering
\begin{threeparttable}
\caption{Zero-field splitting parameters of triplet electronic states measured in GHz.}
\begin{tabular}{|l|c|c|c|}
\hline
 State & Parameter & Theory & Experiment \\
\hline
$(1)^3\bar{A}_2$ & $D$ & 3.27 &  2.88\cite{Loubser-1978} \\
\hline
\hline
\multirow{6}{*}{$(1)^3\bar{E}$} &
$\lambda_z$  & 18.7 & - \\
& $p\lambda_z$  & 4.32 & 5.3\cite{Batalov:PRL2009} \\
\cline{2-4}
& $D_{es}$  & 2.11 & 1.42\cite{Batalov:PRL2009} \\
\cline{2-4}
& $\Delta$  & 2.87 & - \\
& $q \Delta$ & 1.77 & 1.55\cite{Batalov:PRL2009}\\
\hline
\end{tabular}
\label{table:ZFS_data}  
\end{threeparttable}
\end{table}
By introducing SOC and SSC to CASSCF(6e,4o) eigenstates in the framework of quasi-degenerate perturbation theory (QDPT), we computed the fine-structure splitting of the spin-triplet states. The data discussed below (and in section S5 of SI) were calculated at the ground state ($(1)^3\bar{A}_2$) equilibrium geometry, and the MECP geometry between $(1)^3\bar{E}_x$ and $(1)^3\bar{E}_y$ in the case of $(1)^3\bar{E}$. State-specific orbitals were used to express the CASSCF wavefunctions, with 50-50\% weighing in the latter case.  

The $D$ tensor of the $(1)^3\bar{A}_2$ state was obtained as the energy difference between the ground QDPT state (corresponding to $(1)^3\bar{A}_2$, $m_S=0$) and the two lowest excited states ($(1)^3\bar{A}_2$, $m_S\pm1$). We note that spin-spin coupling (SSC) effects dominate in ground-state zero-field splitting - QDPT with only SOC results in three degenerate sublevels for $(1)^3\bar{A}_2$. As shown in the top row of Table~\ref{table:ZFS_data}, a fair reproduction of experimental data ($D$ = 3.27 GHz vs the measured $D$ = 2.88 GHz\cite{Loubser-1978}) was achieved.

Next, we investigated the fine structure of the $(1)^3\bar{E}$ electronic states. In the $C_{3v}$ symmetry of the MECP geometry, two orbitally degenerate states are split into six spin sublevels by SOC and SSC, the relative energy levels of which can be characterized by three parameters: $\lambda_z$, $D_{es}$ and $\Delta$, see Fig. \ref{fig:ZFS} for visual explanation.
The modeling of this splitting is, however, more complicated than simply extracting the raw QDPT energies, as the dynamic Jahn-Teller instability of the $(1)^3\bar{E}$ states attenuates the bare  $\lambda_z$ and $\Delta$ splitting parameters by the so-called Ham reduction factors\cite{Ham_1968}.

The damping of $\lambda_z$ by the dynamic JT effect ($p$ reduction factor) simplifies to $p$ = 0 if $E_{\rm JT} \gg h\nu_e$ holds, and can be expressed as $p = e^{-4E_{\rm JT}/h\nu_e}$ in the opposite extreme case ($E_{\rm JT} \ll h\nu_e$). If the JT stabilization energy and the JT phonon energy are comparable (which is the case for $(1)^3\bar{E}$ according to our calculations), the empirical formula of
\begin{equation*}
p = e^{-1.974(\frac{E_{\rm JT}}{h\nu_e})^{0.761}}   
\end{equation*}
can be used. We note that this formula of Ham assumes pure linear electron-phonon coupling (i.e. Mexican hat potential energy surface with $\delta_{\rm JT}=0$), therefore, we substituted $E_{\rm JT}+\frac{1}{2}\delta_{\rm JT}$ to the numerator of the exponential expression. The damping factor for $\Delta$ ($q$ reduction factor) derives from $p$ as
\begin{equation*}
q = (p+1)/2 \,,  
\end{equation*}
regardless of the range of $E_{\rm JT}/h\nu_e$. In our calculations, the corresponding multiplicators  were found to be $p=0.23$ and $q=0.62$. 

Having the proper values at hand, the damped SOC elements  $p\lambda_z$ and $q\Delta$  can be readily compared to the available experimental data\cite{Batalov:PRL2009} as summarized in  Table~\ref{table:ZFS_data}.
While raw $\lambda_z$ is more than 5 times larger than the experimental value, the Ham-reduced result ($p\lambda_z$ = 4.32 GHz) is in reasonable agreement with the measurements (5.3 GHz). Similarly, after taking into account the $q$ reduction factor, the experimental $\Delta$ splitting of 1.55 GHz is reproduced by our calculations (1.74 GHz) with high accuracy. We note that the ab initio reproduction of these parameters is especially challenging due to the exponentialized $\frac{E_{\rm JT}}{h\nu_e}$ term in the reduction factors, which easily introduces considerable errors. This feature of the ZFS calculations is also manifested in the poor convergence behavior of $p\lambda_z$ with respect to model size (see SI, Table S10), compared to the smooth convergence of all other energies and properties presented in this study. Namely, we found more than 1 GHz difference between the respective values computed for MODEL-3 and MODEL-4. On the other hand, both $\lambda_z$ and the JT parameters required for $p$ (i.e. $E_{\rm JT}$, $\delta_{\rm JT}$, $h\nu_e$) show definite signs of convergence. Therefore, we attribute the apparent model size effect to the unfortunate accumulation of small numerical uncertainties and consider the value of 4.32 GHz as the converged spin-orbit coupling parameter.   

As for the $D_{es}$ parameter, which is unaffected by the DJT effect, we obtained 2.17 GHz, overestimating the experimental value of 1.42 GHz\cite{Batalov:PRL2009} by a factor of 1.5. 

Concerning the deviations of the calculated ZFS parameters from the measured values, an important aspect to note is that the current level of theory lacks the dynamic correlation, as NEVPT2 only corrects energies (and not wavefunctions) in usual implementations of quantum chemical codes. A NEVPT2 correction to the underlying CASSCF wavefunction, if becomes available in the future, is expected to improve the quality of the computational results. It has also not escaped our attention that - given the single-determinant character of $(1)^3\bar{A}_2$ and $(1)^3\bar{E}$ states - the DFT-level computation of ZFS parameters is another possible direction of improvement, albeit we emphasize that it is not generally applicable.

\bigbreak
\section*{\large Discussion}
\label{sec:discussion}
In this theoretical study, we aimed to devise a wave-function-only computational protocol that enables the complete characterization of future quantum bit candidates implemented by point defects in semiconductors. The proposed methodology was tested on the notorious NV$^-$ center in diamond, by reproducing the available experimental data with tolerable error margins.  

Our molecular-cluster-based framework accounts for static electron correlation at CASSCF level, placing the chemically relevant defect orbitals in the active space. Dynamic correlation of the environment was computed at NEVPT2 level which provides a perturbative correction. The cluster geometries were optimized on CASSCF level of theory considering the stiffness of the hosting crystal. We investigated the results for a set of cluster models of increasing size, up to {\textit c.} 300 atoms, in order to confirm the convergence of our predictions. 

We expect that the presented modeling approach can be  routinely  applied  with similar accuracy for color centers  in covalent crystals  featuring  up to 6-10 active defect orbitals.

\section*{\large Methods}
\label{sec:tech}
In this study, we applied the quantum chemical program package ORCA\cite{neese2022software} (version 5.0.3)  for all computations.
For the sake of reproducibility, typical ORCA input files  can be found in the Supporting Information (SI), section S1.1.

The geometries for the conceivable electronic states of NV$^-$ (see Fig.~\ref{fig:Configurations}, bottom) were optimized at state-specific CASSCF(6e,4o)/cc-pVDZ level. In these calculations, we assumed $C_{1h}$ point group as it allows the Jahn-Teller distortion of $\bar{E}$ states ($\bar{A}$ states maintain the $C_{3v}$ symmetry, regardless of the symmetry constraints). Additionally, both the Jahn-Teller barrier and the Jahn-Teller minimum can be located in a straightforward manner by selecting either $A^{\prime}$ or $A^{\prime \prime}$ irreducible representation for the state-specific geometry optimization.

On the obtained equilibrium geometries, SS-CASSCF or SA-CASSCF-NEVPT2 single-point calculations were performed, which provided the energies and properties described in the main text. Varying number of roots were included in the state-average treatment, depending on the purpose of the calculation - see details in the Results section and in the SI.

The CASSCF-NEVPT2 framework was tested on the cluster models using both the strongly contracted (SC) and the fully internally contracted (FIC) implementation of the ORCA suite\cite{Sivalingam2016-pr}, as well as the domain-based local pair natural orbitals (DLPNO) based approximation of the latter\cite{dlpno-nevpt2}. Unfortunately, the computationally less demanding FIC and DLPNO-FIC methods yielded problematic negative eigenvalues for the Koopmans matrix in several cases, even for low-energy excited states. Hence, the  results discussed in the main text were computed using the SC-NEVPT2 approach, as it shows numerical stability and robustness against intruder states.

To speed up calculations, we took advantage of the resolution of identity approximation (RI).  The construction of Coulomb and exchange integrals was carried out using the RIJCOSX\cite{NEESE2009} method with def2/J\cite{Weigend2006} auxiliary basis set. Furthermore, RI-approximated integrals appearing in orbital gradients and Hessians for CASSCF and NEVPT2 were calculated using the cc-pVDZ/C\cite{Weigend2002} auxiliary basis set.

\section*{\large Data availability}
The main data supporting the findings of this study are available within the paper and its Supplementary Information. Further numerical data are available from the authors upon reasonable request.

\section*{\large Acknowledgments} 
We acknowledge fruitful discussions with Rohit Babar, Gergő Thiering, Ádám Gali, Ronald Ulbricht and Michael Roemelt. Helpful computational tweaks learned from the community of the ORCA Forum are also greatly appreciated.  

This research was supported by the National Research, Development, and Innovation Office of Hungary within the Quantum Information National Laboratory of Hungary (Grant No. 2022-2.1.1-NL-2022-00004) and within grants FK 135496 and FK 145395. V.I. also acknowledges the support from the Knut and Alice Wallenberg Foundation through WBSQD2 project (Grant No.\ 2018.0071). Z.B. and A.P.\ acknowledge the financial support of János Bolyai Research Fellowship of the Hungarian Academy of Sciences.

The computations were enabled by resources provided by the National Academic Infrastructure for Supercomputing in Sweden (NAISS) and the Swedish National Infrastructure for Computing
(SNIC) at NSC, partially funded by the Swedish Research Council through grant agreements no. 2022-06725 and no. 2018-05973. We also acknowledge KIF\"U for awarding us computational resources at the Komondor supercomputer in Hungary.

\section*{\large Competing interests}
The authors declare no competing interests.

\section*{\large Author contributions}
Z.B., A.G. and G.B. carried out the calculations and the data analysis. Z.B., G.B. and V.I. wrote the manuscript with inputs from all coauthors. The work was supervised by  G.B. and V.I.

\section*{\large References}
\vskip 0.5cm

\end{document}


\maketitle
\date{
\noindent
$^1$Wigner Research Centre for Physics, PO Box 49, H-1525, Budapest, Hungary\\
$^2$MTA–ELTE Lendület ''Momentum'' NewQubit Research Group, Pázmány Péter sétány 1/A, Budapest, 1117, Hungary\\
$^3$Department of Physics, Chemistry and Biology, Linköping University, SE-581 83 Linköping, Sweden\\
$^4$Department of Atomic Physics, Institute of Physics, Budapest University of Technology and Economics,  M\H{u}egyetem rakpart 3., H-1111 Budapest, Hungary\\
$^5$Department of Physics of Complex Systems, Eötvös Loránd University, Egyetem tér 1-3, H-1053 Budapest, Hungary \\
$^*$email: barcza.gergely@wigner.hu, ivady.viktor@ttk.elte.hu
}


\tableofcontents
\newpage

\section{Theoretical section}

\subsection{Sample input files (ORCA 5.0.3.)}
\label{SM:sect:input}

\subsubsection{Symmetrization of the initial geometry guess}

! RIJCOSX HF cc-pVDZ def2/J noiter xyzfile usesym \\

*xyz -1 3 [Geometry guess (xyz format)] * \\

\# This run recognizes the desired $C_s$ symmetry, cleans up the coordinates and re-orients the geometry.

\subsubsection{Generation of initial HF orbitals}

! RIJCOSX HF cc-pVDZ def2/J usesym \\
\%sym \\
pointgroup ''cs'' \\
end \\

*xyz -1 3 [Symmetrized guess (xyz format)] * 

\subsubsection{Preparing the (6e,4o) active space}

! RIJCOSX HF cc-pVDZ def2/J usesym moread noiter \\
\%moinp ''Initial\_HF\_orbitals.gbw'' \\
\%scf \\
rotate \{orbital1, orbital2\} end \quad \quad \quad \quad \quad \quad  \#Ordering the 4 defect orbitals next to each other by the ''rotate'' feature \\
end \\
\%sym \\
pointgroup ''cs'' \\
end \\

*xyz -1 3 [Symmetrized guess (xyz format)] *

\subsubsection{Geometry optimization (SS-CASSCF)}

! cc-pVDZ moread def2/J usesym opt \\
\%moinp ''Rotated\_HF\_orbitals.gbw'' \\
\%casscf \\
nel 6 \\
norb 4 \\
mult 1 \quad \quad \quad  \#1 for singlet, 3 for triplet states \\
irrep 0 \quad \quad \quad  \#0 for $A'$, 1 for $A''$ states \\
nroots 2 \quad \quad \quad  \#Which state of the specified irrep is optimized (1: lowest-energy state; 2: second lowest state; etc.)  
weights[0]=0,1 \quad \quad \quad  \#Setting the weight of the state of interest to 100\%. \\
maxiter 150 \\ 
printwf det \\
end \\
\%geom \\
Constraints \{C \emph{number\_of\_fixed\_atoms} C\} end \quad \quad \quad  \#Fixed atoms: hydrogens and carbons of the outermost shell\\
end \\
\%sym \\
pointgroup ''cs'' \\ 
end \\

* xyz -1 3 [Symmetrized guess (xyz format)] * \\

\# Note: for MECPs between degenerate $E$ states, 50-50\% weighing was used, as follows: \\

irrep 0,1 \\
nroots 1,2 \quad \quad \quad   \#Which state of the specified irrep is optimized (1: lowest-energy state; 2: second lowest state; etc.)  \\     
bweight 1,1 \quad \quad \quad  \#Setting equal weights for the 2 blocks. \\
weights[0]=1 \quad \quad \quad  \#Setting the weight of the state of interest to 100\% in the first block. \\ 
weights[1]=0,1 \quad \quad \quad  \#Setting the weight of the state of interest to 100\% in the second block. 

\subsubsection{Single-point energy calculation with state-averaging over all electronic states (SA-CASSCF-NEVPT2)}

\# Note: symmetry is not fully supported in ORCA for this procedure. Thus, the energy and properties were computed in C1 symmetry. \\

! rijcosx cc-pVDZ def2/J moread cc-pVDZ/C \\

\%moinp ''CASSCF\_orbitals\_of\_optimized\_geometry.gbw'' \\
\%casscf \\
nel 6 \\
norb 4 \\
mult 3,1 \\
nroots 5,8 \\
printwf det \\
ptmethod sc\_nevpt2 \\
actorbs canonorbs \\
nevpt \\
d3tpre 1e-14 \\
d4tpre 1e-14 \\
end \\
end \\

* xyz -1 3 [Optimized geometry (xyz format)] * \\

\subsubsection{Single-point energy calculation with state-averaging over two electronic states of interest (SA-CASSCF-NEVPT2)}

\# Note: symmetry is not fully supported in ORCA for this procedure. Thus, the energy and properties were computed in C1 symmetry. \\

! rijcosx cc-pVDZ def2/J moread cc-pVDZ/C \\

\%moinp ''CASSCF\_orbitals\_of\_optimized\_geometry.gbw'' \\
\%casscf \\
nel 6 \\
norb 4 \\
mult 3,1 \\
nroots 5,8 \\
printwf det \\
bweight 1,0 \quad \quad \quad  \#Selecting the two states of interest by weighing. $E_x$ and $E_y$ are present with 50-50\% weights. \\  
weights[0]=1,0.5,0.5,0,0 \\
ptmethod sc\_nevpt2 \\
actorbs canonorbs \\
nevpt \\
d3tpre 1e-14 \\
d4tpre 1e-14 \\
selectedroots[0]=1,1,1,0,0 \quad \quad \quad  \#Optional setting; to spare computational time, NEVPT2 can be chosen to be\\ 
selectedroots[1]=0,0,0,0,0,0,0,0 \quad \quad \quad  \# computed only for certain states.\\
end \\
end \\

* xyz -1 3 [Optimized geometry (xyz format)] * \\

\subsubsection{Calculation of SOC and SSC couplings (SS-CASSCF)}

\# Note: symmetry is not fully supported in ORCA for this procedure. Thus, the energy and properties were computed in C1 symmetry. \\

! rijcosx cc-pVDZ def2/J moread cc-pVDZ/C \\
\%moinp ''CASSCF\_orbitals\_of\_optimized\_geometry.gbw'' \\
\%casscf \\
nel 6 \\
norb 4 \\
mult 1 \quad \quad \quad  \#1 for singlet, 3 for triplet states \\
nroots 2 \quad \quad \quad  \#Which state of the specified multiplicity is calculated (1: lowest-energy state; 2: second lowest state; etc.)  
weights[0]=0,1 \quad \quad \quad  \#Setting the weight of the state of interest to 100\%. \\
maxiter 150 \\ 
printwf det \\
\%rel \\
dosoc true \\
dossc true \\
printlevel 4 \\
end \\
end \\

* xyz -1 3 [Optimized geometry (xyz format)] * \\

\subsection{Optimized geometries}

The geometries of molecular models (all relevant electronic states optimized at SS-CASSCF(6e,4o)/cc-pVDZ level) can be found in the supplementary file ''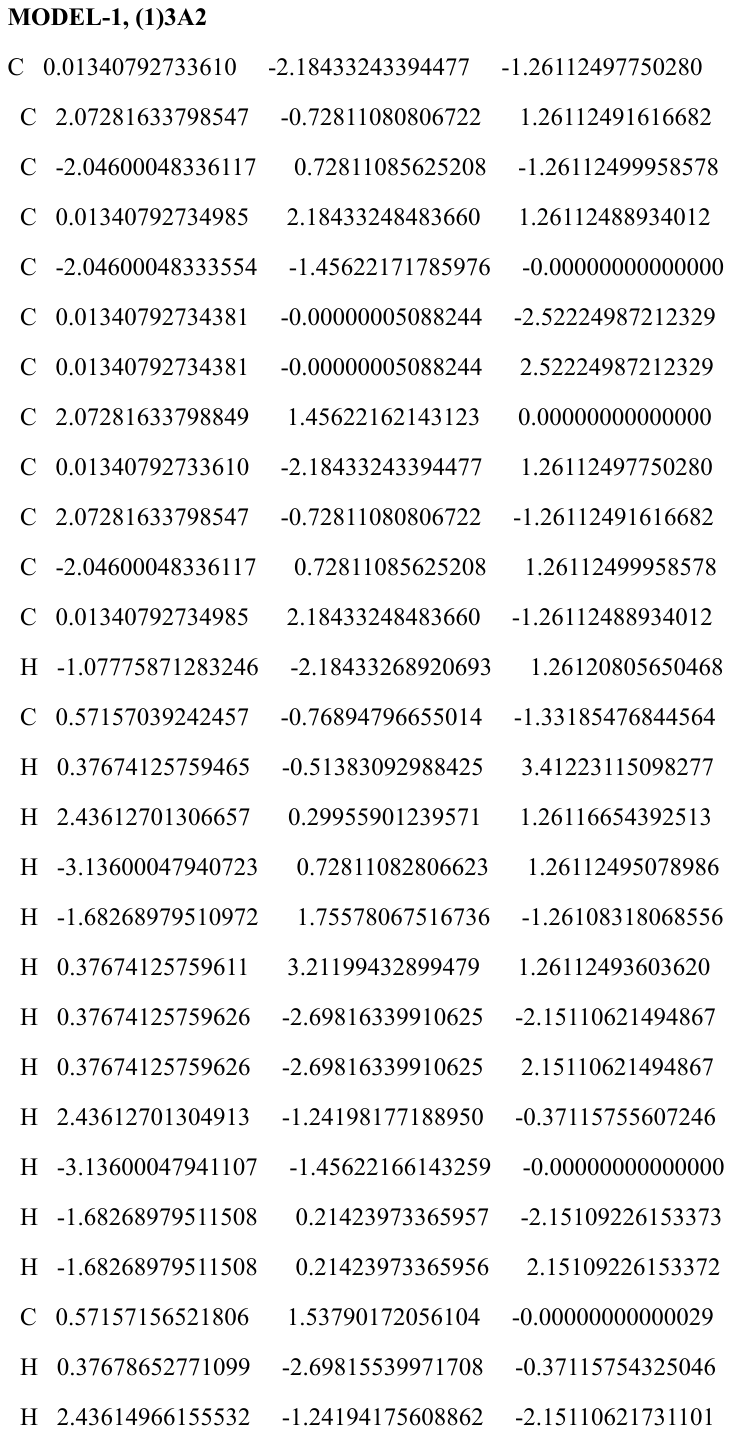''. All xyz coordinates are given in angstroems.

\section{Convergence analysis of the molecular model}

\subsection{Basis size dependence of vertical excitation energies}
\label{SM:sec:basis}

\begin{table}[H]
\setlength\extrarowheight{1pt}
\centering
\scalebox{1}{
\begin{threeparttable}
\begin{tabular}{|l|l|ccccc|}
\hline
State & Model & cc-pVDZ & cc-pVTZ & cc-pVQZ & cc-pV5Z & cc-pV6Z  \\
 \hline
\multirow{4}{*}{$(1)^1\bar{E}$} 
& MODEL-1  & 0.61 (0.90) & 0.58 (0.88) & 0.55 (0.85) & 0.52 (0.83) & 0.50 (0.81)\\
& MODEL-2  & 0.63 (0.96) & 0.61 (0.95) & 0.60 (0.94) & 0.59 (0.94) &  \\
& MODEL-3  & 0.62 (0.97) & 0.60 (0.97) & 0.60 (0.96)\tnote{a} &  &  \\
& MODEL-4  & 0.62 (0.97) & & & &  \\
 \hline
\multirow{4}{*}{$(1)^1\bar{A}_1$} 
& MODEL-1  & 1.83 (3.04) & 1.67 (2.94) &  1.55 (2.85) & 1.43 (2.76) & 1.35 (2.69)\\ 
& MODEL-2  & 1.88 (3.24) & 1.77 (3.20) & 1.71 (3.17) & 1.67 (3.15) &  \\ 
& MODEL-3  & 1.79 (3.27) & 1.71 (3.25) & 1.69 (3.25)\tnote{a} &  &  \\ 
& MODEL-4  &  1.77 (3.28) & & & &  \\
 \hline
\multirow{4}{*}{$(1)^3\bar{E}$} 
& MODEL-1  & 2.27 (3.18) & 2.13 (3.12) & 2.04 (3.05) & 1.96 (2.98) & 1.87 (2.94) \\
& MODEL-2  & 2.37 (3.35) & 2.27 (3.34) & 2.24 (3.32) & 2.22 (3.30) &  \\
& MODEL-3  & 2.34 (3.42) & 2.28 (3.42) & 2.27 (3.42)\tnote{a} &  &  \\
& MODEL-4  &  2.36 (3.43) & & & &  \\
 \hline
\end{tabular}

\begin{tablenotes}
    \item[a]  regarding the computational demands, cc-pVTZ basis set was applied on the  hydrogens
\end{tablenotes}

\end{threeparttable}}
\caption{Basis set dependence of vertical excitation energies (eV) of $(1)^1\bar{E}$, $(1)^1\bar{A}_1$ and $(1)^3\bar{E}$  states with respect to the $(1)^3\bar{A}_2$  ground state for various models at   CASSCF-SC-NEVPT2 level of theory, CASSCF energies are in parenthesis. In the CASSCF calculations, the active space of  $\{a_1,a_1^*,e_x,e_y\}$ defect orbitals  filled with 6 electrons was considered which was optimized for 3-3 roots in the spin triplet and singlet sectors. 
\label{SM:table:vertical_energy_basis_test}  }
\end{table}

\subsection{Model size dependence of vertical excitation energies}
\label{SM:sect:casscf-nevpt2-zpe}

\begin{table*}[h]
\setlength\extrarowheight{2pt}
\centering
\begin{tabular}{|l|cccc|}
\hline
State & MODEL-1 & MODEL-2 & MODEL-3 & MODEL-4 \\
\hline
$(1)^3\bar{A}_2$& 0.00 (0.00)  & 0.00 (0.00) & 0.00 (0.00)& (0.00) (0.00) \\
$(1)^3\bar{E}$  & 2.20 (2.76)  & 2.28 (3.05) & 2.23 (3.10)& 2.21 (3.19)  \\
$(2)^3\bar{E}$  &  5.00 (6.35)  & 4.85 (7.29) & 4.40 (6.86) & 4.00 (7.38)\\
$(1)^3\bar{A}_1$& {\it 7.11*} (9.14)& 6.86 (10.30) & 6.40 (9.97)& 5.74 (10.52) \\
\hline
$(1)^1\bar{E}$  &  0.58 (0.83)  & 0.59 (0.90) & 0.57 (0.92)& 0.55 (0.93) \\
$(1)^1\bar{A}_1$& {\it 1.75*} (2.78)& 1.75 (3.06)& 1.63 (3.08)& 1.55 (3.15)\\
$(2)^1\bar{E}$  & 3.76 (5.18) & 3.68 (6.31) & 3.33 (5.93) & 3.01 (6.58)\\
$(3)^1\bar{E}$  & 5.25 (9.68) & 4.92 (10.14) & 4.05 (10.02)& 3.70 (9.90) \\
$(2)^1\bar{A}_1$& 5.22 (7.86)  & 4.90 (9.00) & 4.53 (8.80)& 4.00 (9.33)\\
$(3)^1\bar{A}_1$& 7.55 (12.09) & 6.70 (12.67)& {\it 5.45*} (12.41)& {\it 5.20*} (12.40)\\
$(4)^1\bar{A}_1$& {\it 7.89*} (18.42)& {\it 4.13*} (17.95)& {\it 4.66*} (17.39)& {\it 5.15*} (16.47)\\
\hline
\end{tabular}
\caption{Model size dependence of vertical energies (eV) at SA-CASSCF(6e,4o)-SC-NEVPT2 level of theory in cc-pVDZ basis (at $^3\bar{A}_2$ geometry). 6 + 10 roots were included in the state-averaging. SA-CASSCF(6e,4o) energies are in paranthesis. Starred values with italic font indicate negative denominators in the NEVPT2 treatment, which questions the reliability of the obtained energy level.}
\label{SM:table:vertical_energies_t6s10}  
\end{table*}

\begin{table*}[h]
\setlength\extrarowheight{2pt}
\centering
\begin{tabular}{|l|cccc|}
\hline
State & MODEL-1 & MODEL-2 & MODEL-3 & MODEL-4 \\
\hline
$(1)^3\bar{A}_2$& 0.00 (0.00) & 0.00 (0.00) & 0.00 (0.00) & 0.00 (0.00) \\
$(1)^3\bar{E}$  & 2.15 (2.86) & 2.24 (3.07) & 2.19 (3.16) & 2.18 (3.16)   \\
$(2)^3\bar{E}$  & 4.79 (6.64) & 4.90 (7.02) & 4.41 (6.90) & 4.38 (6.91) \\
\hline
$(1)^1\bar{E}$  & 0.58 (0.85) & 0.59 (0.91) & 0.56 (0.93) & 0.56 (0.94) \\
$(1)^1\bar{A}_1$& 1.73 (2.83) & 1.75 (3.07) & 1.62 (3.12) & 1.60 (3.15)   \\
$(2)^1\bar{E}$  & 3.48 (5.53) & 3.74 (6.05) & 3.28 (5.99) & 3.32 (6.04) \\
$(3)^1\bar{E}$  & 5.22 (9.79) & 4.92 (10.26)& 4.11 (10.16)& 3.94 (10.17)   \\
$(2)^1\bar{A}_1$& 4.97 (8.21) & 5.09 (8.87) & 4.56 (8.90) & 4.54 (8.96)\\

\hline
\end{tabular}
\caption{Model size dependence of vertical energies (eV) at SA-CASSCF(6e,4o)-SC-NEVPT2 level of theory in cc-pVDZ basis (at $^3\bar{A}_2$ geometry). 5 + 8 roots were included in the state-averaging. SA-CASSCF(6e,4o) energies are in paranthesis.}
\label{SM:table:vertical_energies}  
\end{table*}

\newpage

\subsection{Defect orbitals}

\subsubsection{Löwdin orbital composition}
A general an overview of the spatial expansion of orbitals can be obtained by the Löwdin orbital composition, We performed the analysis focusing on the four defect orbitals obtained at CASSCF(6,4) level of theory for all the models. The raw weights printed by ORCA with 0.1~\% precision were adjusted to retrieve 100~\% in total. 
In this study, we integrated the atomic contributions for each atomic coordination sphere  considering the terminating hydrogens as well which form the outermost shell of atoms in the models.
Thus, in total,  MODEL-1, 2, 3, and 4 consists of  3, 4, 5, and 6 atomic shells around the vacancy respectively.
\begin{figure*}[!h]
\begin{center}
	\includegraphics[width=0.92\textwidth]{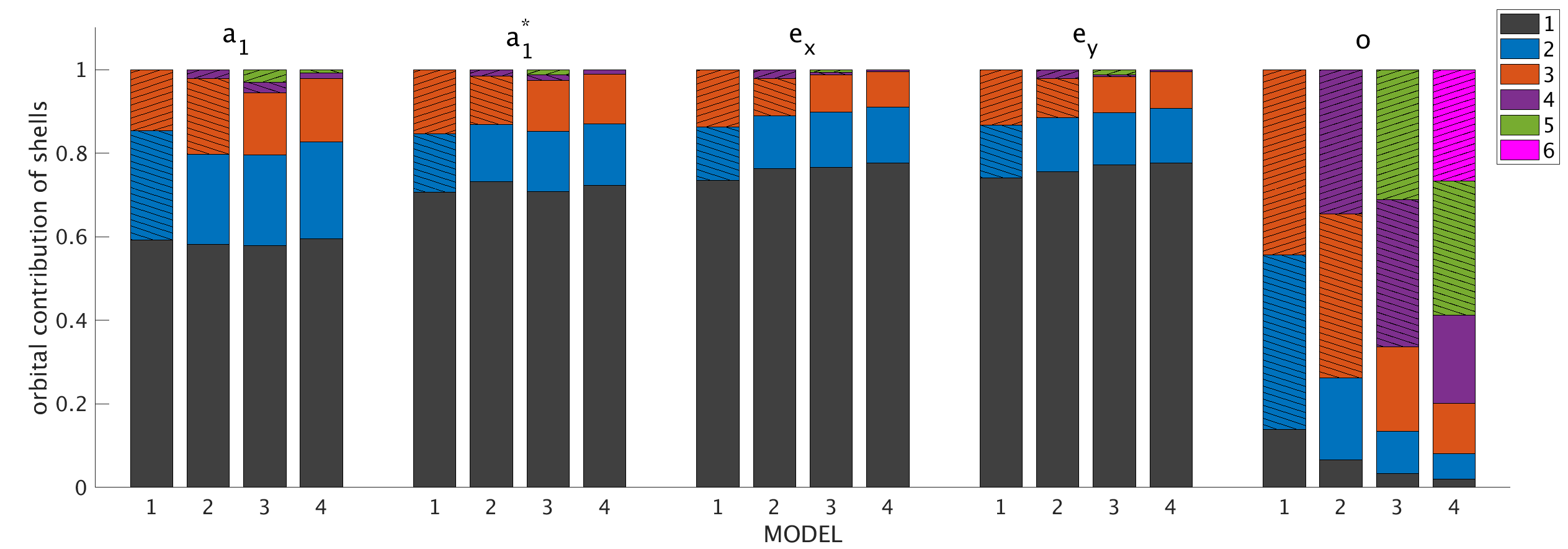}
	\caption{Shellwise Löwdin orbital composition of the   defect orbitals $a_1$, $a_1^*$, $\bar{E}_x$,  $\bar{E}_y$ and the average orbital  $o$ for MODEL-1, 2, 3 and 4 at SA-CASSCF(6,4)/cc-pVDZ level.  For better visibility, the constrained atomic shells $n+1$ and $n+2$ of MODEL-$n$, which are formed of the outermost carbons and the terminating hydrogens,  are distinguished from the relaxed shells by the hatch patterns.}
	\label{SM:fig:loewdin}  
 \end{center}
\end{figure*}

The results for the models were summarized in Fig.~\ref{SM:fig:loewdin}. The defect orbitals $e_x$ and $e_y$  filled with 1.0-1.0 electrons at CASSCF level give the immobilized radical character of the triplet ground state. Correspondingly, they are predominantly localized on the four first-shell atoms with a weight of 73-78\%.
Note that in MODEL-3 and 4, formed of 5 and 6 coordination spheres  respectively, only the three inner shells have significant contribution while the outer atoms have a weight of around 1.5-0.5\% in total. Most importantly, the orbitals have minimal contribution from the  outermost carbon and hydrogen shells which  which  merely  serve to terminate properly the molecules modeling the crystal defect.  

The $a_1^*$ defect orbital has a shellwise Löwdin composition which is rather similar to the counterpart  of the $e$ orbital pair  while the  $a_1$  reveals slightly different behavior. In particular, the weight of the first shell is only around 58\% and the second coordination sphere   has a more pronounced contribution (22-26\%)  than in case of  $a_1^*$ and the $e$ pair (12-14\%). Nevertheless, in MODEL-3 and 4, similarly to the other defect orbitals, the total weight of the outermost shells in the Löwdin composition is insignificant for  $a_1$.  

As a comparison, we also presented the composition of the average  orbital delocalized over the complete model structure which reflects the number of atomic orbitals in the distinct shells. It it remarkable that the innermost shell of four atoms has a marginal  2~\% weight in MODEL-4.

In summary, we found that all defect orbitals are localized based on the Löwdin composition. In MODEL-3 and 4, the defect orbitals expand only in the inner three shells of the structure (up to 95-99~\%) implying that these models are expected to be  sizeable enough to describe the NV center in diamond  properly.

\subsubsection{Radial probability distribution}
\label{SM:sect:radial_distrib}

Regarding the largely different character and physical extension of the atomic orbitals with distinct quantum numbers, the Löwdin molecular orbital composition provides only a rough overview. Hence, we also studied the spacial probability distribution of the defect orbitals to gain further insight. 

\begin{figure*}[!h]
\begin{center}
\includegraphics[width=1\textwidth]{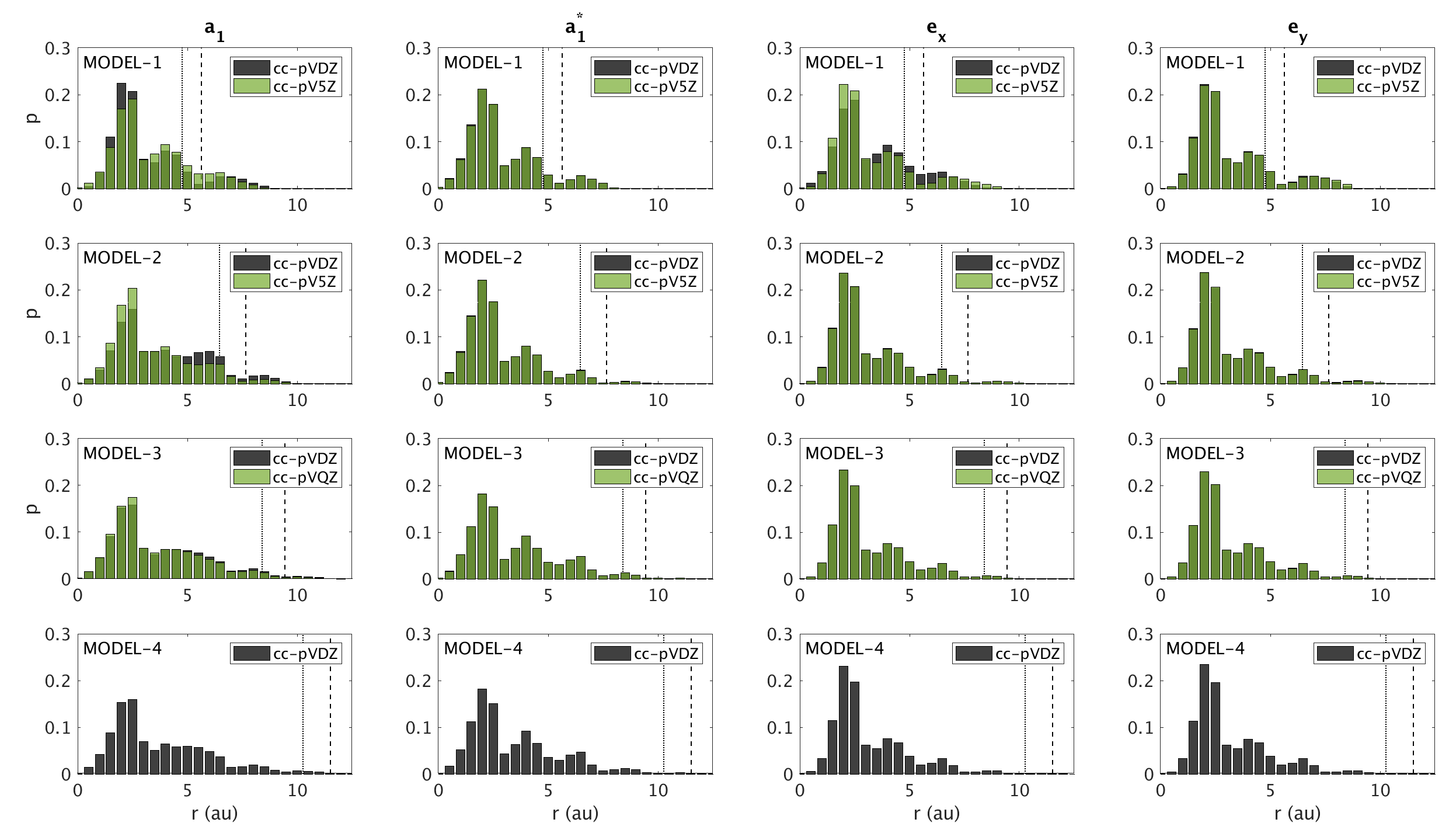}
\caption{Radial distribution of the defect-orbital electronic density. Data obtained for defect orbitals  $a_1$, $a_1^*$, $\bar{E}_x$,  $\bar{E}_y$  is presented for MODEL-1, 2, 3, 4. The distances are  measured from the vacancy site (0,0,0) in atomic unit. The black dotted and dashed lines denote the average distance of the atoms in outermost carbon shell  and the hydrogen atoms respectively.  Note that regarding the computational demand of MODEL-3,  hydrogens were represented by the cc-pVTZ basis set in cc-pVQZ basis calculation similarly to Table~\ref{SM:table:vertical_energy_basis_test}.}
	\label{SM:fig:radial_dist}  
 \end{center}
\end{figure*}

The radial probabilities, measured from the vacancy site at the origin, were obtained from the corresponding cube files generated from the ORCA output data. State averaged CASSCF ground-state orbital data were demonstrated for various basis sets in order to analyze the convergence of the defect orbitals. Results were collected in Fig.~\ref{SM:fig:radial_dist}.

For MODEL-1, we found that all four defect orbitals have substantial  radial weight in the  vicinity and well beyond the terminating hydrogens (see their averaged distance denoted by the dashed vertical line). While the $a_1^*$ orbital is practically insensitive to the basis set, the $e_x$ and  $a_1$ molecular orbitals depend more pronounced on the underlying atomic orbitals. Most notably,  defect orbital $a_1$ show a tendency to gain significant weight at around the hydrogens at the expense of loosing contribution close to the defect center in the larger basis set (cc-pV5Z).
 
For  gradually increasing model sizes, we found that all defect orbitals have weaker basis dependence and they have  decreasing weight in the vicinity of the corresponding hydrogen shell. In fact, the radial distributions for MODEL-3 and 4  not only look practically alike but also have already negligible weight at the artifact terminating coordination spheres.

\newpage

\section{Model size dependence of relaxed excitation energies and molecular geometries }
\label{SM:sect:casscf-nevpt2-zpe}

\subsection{Range of geometry relaxation effects}

In order to quantify the effect of the geometry optimization  in the distinct atomic shells, we analyzed the root mean square deviation (RMSD) of the atomic positions of the ground state clusters  from the unrelaxed structure, see Table~\ref{SM:table:RMSD}.
We found that the relaxation is the largest  in the first shell (0.1 {\AA}), and they decay fast with distance from the defect (0.03-0.01 {\AA} in shells 2 and 3).
Interestingly, for MODEL-4 the average relaxation effect in shells 3 and 4 was found to be comparable contrary to the naive expectation, i.e., it is in the range of 0.01 {\AA} in both cases. The observed faint variance of RMSD in these coordination spheres might indicate the limitation of the used CASSCF relaxation approach on the outer shells of large models. Nevertheless, the potential inaccuracy of the applied approximation is not only  marginal but more importantly any consequent error is also expected to be cancelled out in the relative energy of the investigated excitations, which are governed by the localized defect orbitals.

We also studied the similarity of the relaxed cluster structures of  different sizes by the RMSD between  MODEL-4 and the smaller models. While the optimized shells of MODEL-1 and MODEL-2 have deviations of 0.012-0.025 {\AA} from MODEL-4, in case of MODEL-3, the corresponding RMSD values are only 0.002-0.004 \AA. 

The observed similarity of the relaxed MODEL-3 and MODEL-4 ground state structures also implies that these clusters can be  considered large enough to provide a convergent molecular model of the NV$^-$ defect center.

\begin{table}[H]
\setlength\extrarowheight{0pt}
\centering
\scalebox{0.9}{
\begin{tabular}{|c|ccc||cccc|}
\hline
 & \multicolumn{3}{c||}{reference: MODEL-4} &\multicolumn{4}{c|}{reference: unrelaxed structure} \\
\hline
shells & MODEL-1 & MODEL-2 & MODEL-3 & MODEL-1 & MODEL-2 & MODEL-3 & MODEL-4 \\
 \hline
1   &  0.0247 & 0.0169  &  0.0019 & 0.1105 &0.0979 & 0.1117 & 0.1127\\
2   & 0.0284 & 0.0116 &   0.0018 & - & 0.0225 &0.0271 &0.0284\\
3   &  - &  0.0123    &0.0037 & - & - & 0.0095 & 0.0123 \\
4  & - & - & 0.0103 & - & - & - & 0.0103\\
 \hline
\end{tabular}}
\caption{Root mean square deviation (\AA) of the position of the second-row atoms in the ground state  of the  MODELs from counterparts of MODEL-4 and  the unrelaxed geometry  for the distinct atomic shells. Relaxations were performed at SS-CAS(6,4)/ccpVDZ level of theory. Recall, in MODEL-$n$, the inner $n$ shells  are relaxed while $n+1$ and $n+2$ shells  are kept at fixed position corresponding to the hosting diamond structure. 
\label{SM:table:RMSD}  }
\end{table}

\subsection{Convergence of geometry: C-C distances in the innermost shell}

\begin{table*}[h]
\setlength\extrarowheight{1pt}
\centering
\resizebox{0.6\textwidth}{!}{%
\begin{tabular}{|l|c|c|c|c|}
\hline
 Relaxed state &  MODEL-1 & MODEL-2 & MODEL-3 & MODEL-4  \\
\hline
$(1)^3\bar{A}_2$ & 2.66 (x3) & 2.63 (x3) & 2.63 (x3) & 2.63 (x3)  \\
$(1)^3\bar{E}_x\rightarrow{}^3\bar{A}''$ & 2.79 (x2), 2.73 & 2.77 (x2), 2.70 & 2.79 (x2), 2.70 & 2.80 (x2), 2.70 \\
$(1)^3\bar{E}_y\rightarrow{}^3\bar{A}'$  & 2.77 (x2), 2.79 & 2.74 (x2), 2.78 & 2.76 (x2), 2.79 & 2.77 (x2), 2.80 \\
$(2)^3\bar{E}_x\rightarrow{}^3\bar{A}''$  & 2.74 (x2), 2.67 & 2.72 (x2), 2.64 & 2.74 (x2), 2.64 & 2.75 (x2), 2.63 \\
$(2)^3\bar{E}_y\rightarrow{}^3\bar{A}'$  & 2.73 (x2), 2.73 & 2.70 (x2), 2.72 & 2.72 (x2), 2.74 & n/a \\
\hline
$(1)^1\bar{E}_x\rightarrow{}^1\bar{A}''$  & 2.73 (x2), 2.61 & 2.70 (x2), 2.58 & 2.71 (x2), 2.57 & 2.72 (x2), 2.57 \\
$(1)^1\bar{E}_y\rightarrow{}^1\bar{A}'$  & 2.69 (x2), 2.74 & 2.67 (x2), 2.70 & 2.68 (x2), 2.70 & 2.68 (x2), 2.70  \\
$(1)^1\bar{A}_1$  & 2.72 (x3) & 2.69 (x3) & 2.70 (x3) & 2.70 (x3) \\
$(2)^1\bar{E}_x\rightarrow{}^1\bar{A}''$ & 2.68 (x2), 2.61 & 2.66 (x2), 2.57& 2.68 (x2), 2.56 &  2.68 (x2), 2.56\\
$(2)^1\bar{E}_y\rightarrow{}^1\bar{A}'$  & n/a & 2.66 (x2), 2.65 & n/a & n/a \\
$(3)^1\bar{E}_x\rightarrow{}^1\bar{A}''$  & 2.77 (x2), 2.79 & 2.71 (x2), 2.64 & 2.72 (x2), 2.66 & 2.74 (x2), 2.67 \\
$(3)^1\bar{E}_y\rightarrow{}^1\bar{A}'$  & n/a & n/a & n/a & n/a \\
$(2)^1\bar{A}_1$ & 2.76 (x3) & 2.74 (x3) & 2.75 (x3) & 2.76 (x3) \\
\hline
\end{tabular}}
\caption{Carbon-carbon interatomic distances (\AA) in the innermost shell of atoms. n/a: Optimization failed due to technical issues (uncontrollable flipping of CASSCF roots).}
\label{SM:table:distances}  
\end{table*}

\newpage
\subsection{Calculation of relaxed excitation energies}

The energy levels relative to the ground state ($(1)^3\bar{A}_2$) presented in Fig. 4 of the main text were determined as follows. 

In the first step, energy differences between $C_{3v}$ symmetrical structures were calculated, using NEVPT2 on top of a CASSCF(6e,4o) wavefunction, which was optimized for two selected states. (If an $\bar{E}$ state was involved, the mixture of 50\% $\bar{E}_x$ and 50\% $\bar{E}_y$ was considered as one of the states.) To gain as accurate energies as possible, such two-state calculations were performed for energetically neighboring states. That is, the ground state and the first excited state, then the first and second excited states, etc., were compared in this manner for each multiplicity block. 

While the geometry of $\bar{A}$ states has $C_{3v}$ symmetry by construction, additional geometry optimization is required for $\bar{E}$ states to gain a $C_{3v}$ symmetrical energy minimum, for which the NEVPT2 energy difference can be reliably computed. In the case of $(1)^3\bar{E}$ and $(1)^1\bar{E}$, we managed to locate the minimum-energy crossing point (MECP) of $\bar{E}_x$ and $\bar{E}_y$ by geometry optimization with 50-50\% weighing, and the relaxation within $C_{3v}$ symmetry constraint was studied on these structures. For $(2)^3\bar{E}, (2)^1\bar{E}$ and $(3)^1\bar{E}$, on the other hand, the MECP optimization failed, and we needed to use the $C_{3v}$ symmetrical geometry of other electronic states to approximate the MECP. 

Tab. \ref{SM:table:calc1} summarizes the results of two-state CASSCF-NEVPT2 calculations, which altogether describe the relative energy levels at $C_{3v}$ symmetry.

\begin{table*}[h]
\setlength\extrarowheight{1pt}
\centering
\resizebox{\textwidth}{!}{%
\begin{tabular}{|c|c|c|c|c|c|c|c|c|}
\hline
No. & Transition & Initial geom. & Final geom. & Weighing & MODEL-1 & MODEL-2 & MODEL-3 & MODEL-4  \\
\hline
1& $(1)^3\bar{A}_2\rightarrow(1)^3\bar{E}\phantom{_1}$ & $(1)^3\bar{A}_2$ & $(1)^3\bar{E}$ MECP & 50\% $(1)^3\bar{A}_2$, 25\%$(1)^3\bar{E}_x$, 25\%$(1)^3\bar{E}_y$ &  2.01 eV & 1.99 eV & 1.93 eV & 1.89 eV  \\
\hline
2& $(1)^3\bar{E}\phantom{_1}\rightarrow(2)^3\bar{E}\phantom{_1}$ & $(1)^3\bar{E}$ MECP & $(1)^1\bar{A}_1$* & 25\%$(1)^3\bar{E}_x$, 25\%$(1)^3\bar{E}_y$, 25\%$(2)^3\bar{E}_x$, 25\%$(2)^3\bar{E}_y$ &  3.11 eV & 3.26 eV & 2.90 eV & 2.93 eV  \\
\hline
3& $(1)^3\bar{A}_2\rightarrow(1)^1\bar{E}\phantom{_1}$ & $(1)^3\bar{A}_2$ & $(1)^1\bar{E}$ MECP & 50\% $(1)^3\bar{A}_2$, 25\%$(1)^1\bar{E}_x$, 25\%$(1)^1\bar{E}_y$ &  0.57 eV & 0.59 eV & 0.57 eV & 0.56 eV  \\
\hline
4& $(1)^1\bar{E}\phantom{_1}\rightarrow(1)^1\bar{A}_1$ & $(1)^1\bar{E}$ MECP & $(1)^1\bar{A}_1$\phantom{*} & 25\%$(1)^1\bar{E}_x$, 25\%$(1)^1\bar{E}_y$, 50\% $(1)^1\bar{A}_1$ &  1.22 eV & 1.23 eV & 1.15 eV & 1.09 eV  \\
\hline
5& $(1)^1\bar{A}_1\rightarrow(2)^1\bar{E}\phantom{_1}$ & $(1)^1\bar{A}_1$ & $(1)^3\bar{A}_2$* & 50\% $(1)^1\bar{A}_1$, 25\%$(2)^1\bar{E}_x$, 25\%$(2)^1\bar{E}_y$ & 1.98 eV & 2.31 eV & 2.01 eV & 1.89 eV  \\
\hline
6& $(1)^1\bar{A}_1\rightarrow(2)^1\bar{A}_1$ & $(1)^1\bar{A}_1$ & $(2)^1\bar{A}_1$\phantom{*} & 50\% $(1)^1\bar{A}_1$, 50\%$(2)^1\bar{A}_1$ & 3.84 eV & 2.84 eV & 2.50 eV & 2.48 eV  \\
\hline
7& $(2)^1\bar{A}_1\rightarrow(3)^1\bar{E}\phantom{_1}$ & $(2)^1\bar{A}_1$ & $(1)^1\bar{A}_1$* & 50\% $(2)^1\bar{A}_1$, 25\%$(3)^1\bar{E}_x$, 25\%$(3)^1\bar{E}_y$ & -0.25 eV & 0.54 eV & 0.22 eV & 0.24 eV  \\
\hline
\end{tabular}}
\caption{Relaxation among $C_{3v}$ symmetrical geometries. Energy differences were determined at SA-CASSCF(6e,4o)-NEVPT2 level, with the indicated weighings in orbital optimization. * The optimization of MECP failed. Instead, the geometry of another $C_{3v}$ symmetrical electronic state was used to study the relaxation within $C_{3v}$ symmetry. The indicated state was chosen based on geometrical similarity to the known JT minimum of the final state.}
\label{SM:table:calc1}  
\end{table*}

As the next step, the energy contribution of Jahn-Teller distortion of $E$ states was calculated. As the JT relaxation proceeds within a single potential energy surface, state-specific CASSCF calculations are sufficient to describe this effect. The SS-CASSCF results are shown in Tab. \ref{SM:table:calc2}.

\begin{table*}[h]
\setlength\extrarowheight{1pt}
\centering
\resizebox{\textwidth}{!}{%
\begin{tabular}{|c|c|c|c|c|c|c|c|c|}
\hline
No. & Transition & Initial geom. & Final geom. & Weighing & MODEL-1 & MODEL-2 & MODEL-3 & MODEL-4  \\
\hline
8& $(1)^3\bar{E}\rightarrow{}^3\bar{A}''$ & $(1)^3\bar{E}$ MECP& $(1)^3\bar{E}_x\rightarrow{}^3\bar{A}''$ JT-min. & 100\% $(1)^3\bar{E}_x\rightarrow{}^3\bar{A}''$ &  -0.03 eV & -0.05 eV & -0.07 eV & -0.07 eV  \\
\hline
9& $(1)^3\bar{E}\rightarrow{}^3\bar{A}'\phantom{'}$ & $(1)^3\bar{E}$ MECP& $(1)^3\bar{E}_y\rightarrow{}^3\bar{A}'\phantom{'}$ JT-min. & 100\% $(1)^3\bar{E}_y\rightarrow{}^3\bar{A}'\phantom{'}$ &  -0.02 eV & -0.03 eV & -0.04 eV & -0.04 eV  \\
\hline
10& $(2)^3\bar{E}\rightarrow{}^3\bar{A}''$ & $(1)^1\bar{A}_1$ & $(2)^3\bar{E}_x\rightarrow{}^3\bar{A}''$ JT-min. & 100\% $(2)^3\bar{E}_x\rightarrow{}^3\bar{A}''$ &  -0.33 eV & -0.32 eV & -0.37 eV & -0.38 eV  \\
\hline
11& $(1)^1\bar{E}\rightarrow{}^1\bar{A}''$ & $(1)^1\bar{E}$ MECP& $(1)^1\bar{E}_x\rightarrow{}^1\bar{A}''$ JT-min. & 100\% $(1)^1\bar{E}_x\rightarrow{}^1\bar{A}''$ &  -0.11 eV & -0.12 eV & -0.14 eV & -0.15 eV  \\
\hline
12& $(1)^1\bar{E}\rightarrow{}^1\bar{A}'\phantom{'}$ & $(1)^1\bar{E}$ MECP& $(1)^1\bar{E}_y\rightarrow{}^1\bar{A}'\phantom{'}$ JT-min. & 100\% $(1)^1\bar{E}_y\rightarrow{}^1\bar{A}'\phantom{'}$ &  -0.07 eV & -0.07 eV & -0.13 eV & -0.12 eV  \\
\hline
13& $(2)^1\bar{E}\rightarrow{}^1\bar{A}''$ & $(1)^3\bar{A}_2$ & $(2)^1\bar{E}_x\rightarrow{}^1\bar{A}''$ JT-min. & 100\% $(2)^1\bar{E}_x\rightarrow{}^1\bar{A}''$ &  -0.18 eV & -0.25 eV & -0.30 eV & -0.32 eV  \\
\hline
14& $(3)^1\bar{E}\rightarrow{}^1\bar{A}''$ & $(1)^1\bar{A}_1$ & $(3)^1\bar{E}_x\rightarrow{}^1\bar{A}''$ JT-min. & 100\% $(3)^1\bar{E}_x\rightarrow{}^1\bar{A}''$ &  -0.23 eV & -0.32 eV & -0.18 eV & -0.20 eV  \\
\hline
\end{tabular}}
\caption{Relaxation from $C_{3v}$ symmetrical geometries to Jahn-Teller minima. Energy differences were determined at SS-CASSCF(6e,4o) level.}
\label{SM:table:calc2}  
\end{table*}

Finally, the relaxation energies were combined based on Hess's law to obtain comparable energy levels for all states, as presented in Tab. \ref{SM:table:calc3}. 

\begin{table*}[h]
\setlength\extrarowheight{1pt}
\centering
\begin{tabular}{|c|c|c|c|c|c|}
\hline
State & Summed energies (Tabs. S6-S7) & MODEL-1 & MODEL-2 & MODEL-3 & MODEL-4  \\
\hline
$(1)^3\bar{A}_2$ & - & 0.00 eV & 0.00 eV & 0.00 eV & 0.00 eV \\
\hline
$(1)^3\bar{E}_x\rightarrow{}^3\bar{A}''$ & 1 + 8 & 1.98 eV & 1.94 eV & 1.86 eV & 1.81 eV \\
\hline
$(1)^3\bar{E}_y\rightarrow{}^3\bar{A}'\phantom{'}$ & 1 + 9 & 2.00 eV & 1.97 eV & 1.89 eV & 1.84 eV \\
\hline
$(2)^3\bar{E}_x\rightarrow{}^3\bar{A}''$ & 1 + 2 + 10 & 4.80 eV & 4.93 eV & 4.47 eV & 4.44 eV \\
\hline
$(1)^1\bar{E}_x\rightarrow{}^1\bar{A}''$ & 3 + 11 & 0.47 eV & 0.47 eV & 0.42 eV & 0.41 eV \\
\hline
$(1)^1\bar{E}_y\rightarrow{}^1\bar{A}'\phantom{'}$ & 3 + 12 & 0.50 eV & 0.51 eV & 0.44 eV & 0.44 eV \\
\hline
$(1)^1\bar{A}_1$ & 3 + 4 & 1.79 eV & 1.81 eV & 1.72 eV & 1.65 eV \\
\hline
$(2)^1\bar{E}_x\rightarrow{}^1\bar{A}''$ & 3 + 4 + 5 + 13 & 3.58 eV & 3.87 eV & 3.43 eV & 3.21 eV \\
\hline
$(2)^1\bar{A}_1$ & 3 + 4 + 6 & 5.62 eV & 4.65 eV & 4.22 eV & 4.12 eV \\
\hline
$(3)^1\bar{E}_x\rightarrow{}^1\bar{A}''$ & 3 + 4 + 6 + 7 + 14 & 5.14 eV & 4.86 eV & 4.27 eV & 4.17 eV \\
\hline

\end{tabular}
\caption{Combination of energy contributions presented in Tabs. \ref{SM:table:calc1}-\ref{SM:table:calc2} to energy levels relative to the ground electronic state.}
\label{SM:table:calc3}  
\end{table*}

\newpage

\newpage
\mbox{}
\newpage

\section{ Model size dependence of Jahn-Teller parameters }
\label{SM:sect:JT}

\subsection{Results of linear PES scans (calculation of $h\nu_E$)}

The energy of the effective Jahn-Teller phonon ($h\nu_E$) was determined as follows.

\begin{figure}[H]
\begin{center}
 \includegraphics[width=0.95\textwidth]{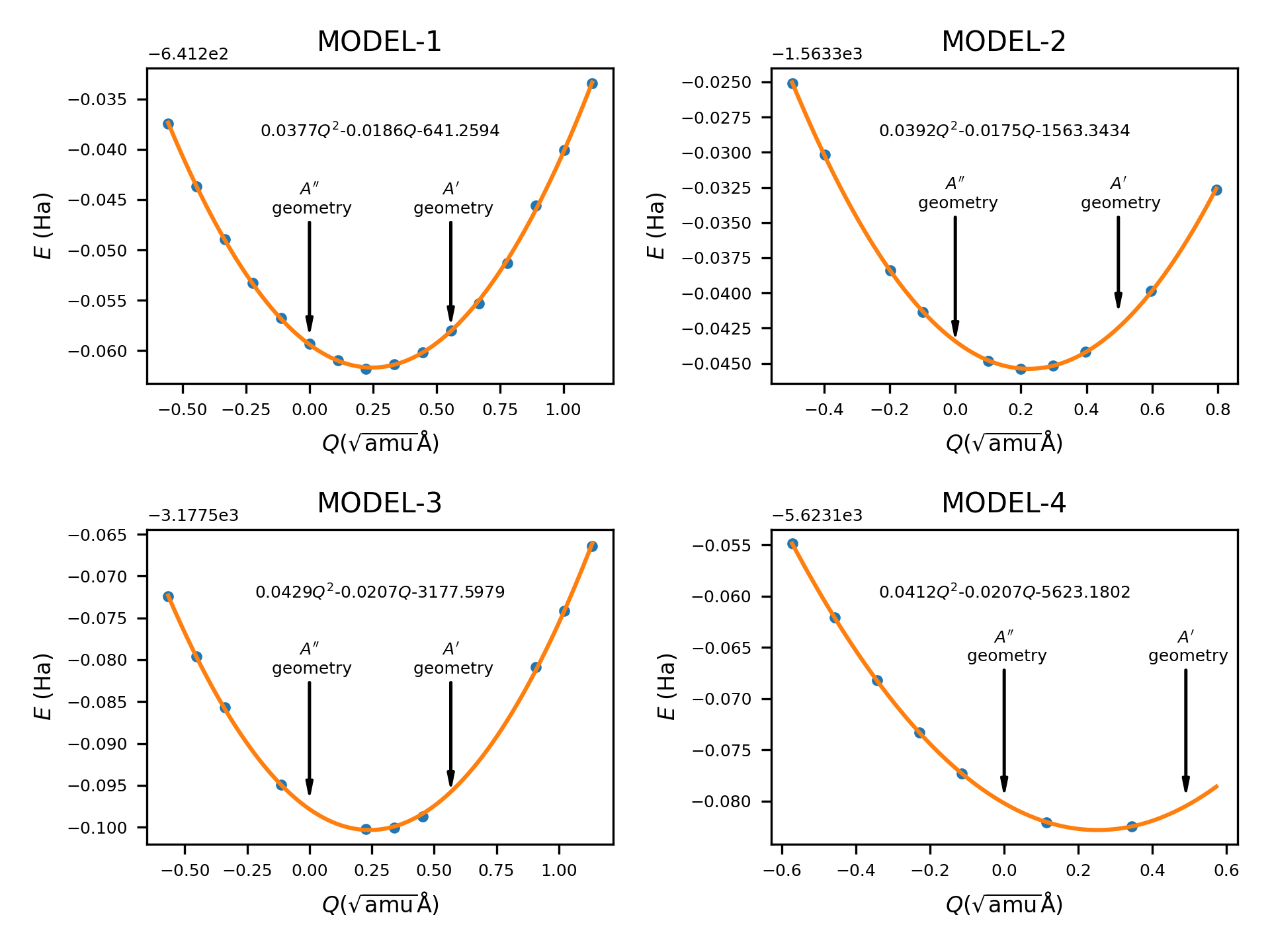}
	\caption{Linear PES scan between $\bar{A}'$ and $\bar{A}''$ equilibrium geometries of the $(1)^3\bar{E}$ electronic state.}
\end{center}
	\label{SM:fig:JT1}  
\end{figure}

\begin{figure}[H]
\begin{center}
 \includegraphics[width=0.95\textwidth]{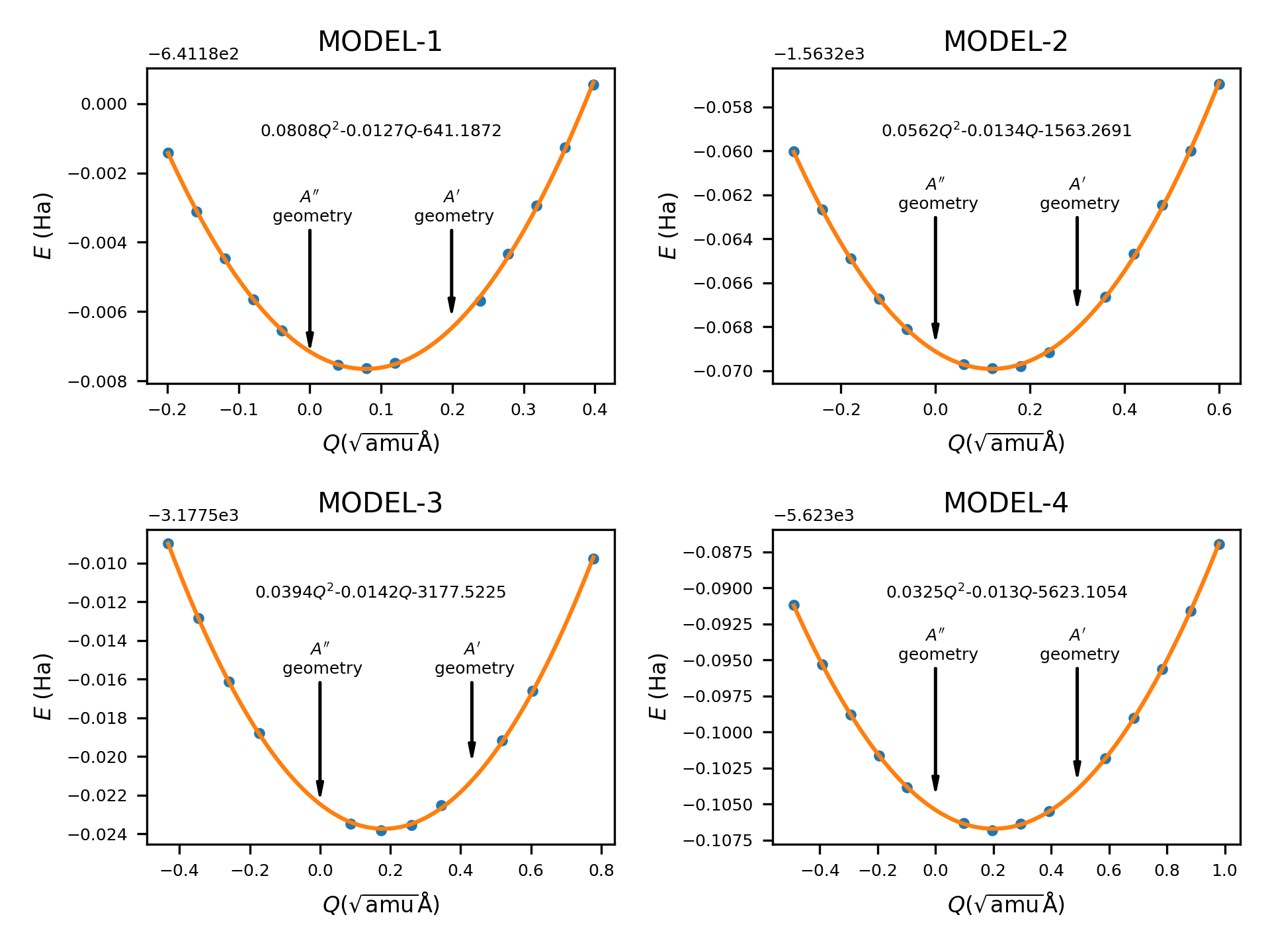}
	\caption{Linear PES scan between $\bar{A}'$ and $\bar{A}''$ equilibrium geometries of the $(1)^1\bar{E}$ electronic state.}
	\label{SM:fig:JT2}  
\end{center}
\end{figure}

In the first step, several geometries were generated with linear interpolation/extrapolation, based on the coordinates of $\bar{A}'$ and $\bar{A}''$ equilibrium geometries. Then, the CASSCF energy was computed for all geometries, using 50-50\% weighing for $\bar{A}'$ and $\bar{A}''$ representations of the electronic state of interest. (This weighing gives minimal energy at $C_{3v}$ symmetry, thus mimics the absence of JT effect.) The results of these calculations are summarized in Figs.~\ref{SM:fig:JT1}-\ref{SM:fig:JT2}. Herein, the vertical axis shows the CASSCF energy, while the horizontal axis represents the deviation of the geometry from $\bar{A}''$ equilibrium, given in a single mass-weighted coordinate (Q). The latter derives from the mass-weighted coordinates of individual atoms as 
\begin{equation}
Q = \sqrt{\sum m_ir^2_i}     
\end{equation}
where  $m_i$ is the mass of the $i$th atom and $r_i$ is the distance of this atom from its position in the reference geometry (i.e. $\bar{A}''$ equilibrium). We note that $r_i$ equals 0 for the outermost shell of atoms, as they are fixed during geometry optimization (hence also during the linear PES can).
Assuming harmonic nuclear movement, the electronic energy depends on Q as
\begin{equation}
E = \frac{1}{2}\omega^2(Q-Q_{min})^2    
\end{equation}
where $Q_{min}$ belongs to the atomic positions of the energy minimum and $\omega$ is the angular frequency of the JT vibration. Thus, $\omega^2$ can be extracted from the quadratic coefficient fitted to the parabolas shown in Figs. \ref{SM:fig:JT1}-\ref{SM:fig:JT2}. We note that the unit conversion from $Ha/amu\AA^2$ to the familiar $1/s^2$ can be achieved by a multiplication by $2.63\times10^{29}$.

Having $\omega$ at hand, the phonon energies shown in Tab. \ref{SM:table:JT} can be simply obtained as $\hbar\omega$.

\subsection{Overview of Jahn-Teller parameters }

\begin{table*}[h]
\setlength\extrarowheight{1pt}
\centering
\begin{tabular}{|l|l|cccc|}
\hline
State& Parameter &  MODEL-1 & MODEL-2 & MODEL-3 & MODEL-4  \\
\hline
\multirow{5}{*}{$(1)^3\bar{E}$} 
& $d$ (\AA) & 2.78  & 2.76 & 2.77 & 2.78 \\
& $E_{\rm JT}(\bar{A}'')$ (meV) & 35.0  & 52.8  & 67.4  & 72.4 \\
& $E_{\rm JT}(\bar{A}')$ (meV) & 17.0  & 28.3  & 41.1  & 43.0 \\
& $E_{\rm JT}(avg.)$ (meV) & 26.0  & 40.6  & 54.3  & 57.7 \\
& $\delta_{\rm JT}$ (meV) & 18.0 & 24.5 & 26.3 & 29.3 \\
& $h\nu_e$ (meV) & 135.5  & 112.6  & 91.3 & 85.8  \\
\hline
\multirow{5}{*}{$(1)^1\bar{E}$} 
& $d$ (\AA) & 2.70 & 2.66 & 2.67 & 2.67 \\
& $E_{\rm JT}(\bar{A}'')$ (meV) & 108.9  & 115.5  & 143.1  &  146.6 \\
& $E_{\rm JT}(\bar{A}')$ (meV) & 73.3  & 72.6  & 102.5  &  104.1 \\
& $E_{\rm JT}(avg.)$ (meV) & 91.1  & 94.1  & 122.8 &  125.4 \\
& $\delta_{\rm JT}$ (meV)& 35.4  & 43.2  & 40.6  & 42.5  \\
& $h\nu_e$ (meV) & 92.0  & 94.3  & 98.8  & 96.7  \\
\hline
\end{tabular}
\caption{Model-size dependence of key Jahn-Teller parameters: carbon-carbon interatomic distances in the innermost shell of atoms in MECP ($d$); Jahn-Teller barrier ($\delta_{\rm JT}$); Jahn-Teller stabilization energy ($E_{\rm JT}$); energy of the phonon driving the JT distortion ($h\nu_e$). Note that the energy minimum was found in the $\bar{A}''$ irreducible representation in all  cases.}
\label{SM:table:JT}  
\end{table*}

\newpage

\section{Model size dependence of zero-field splitting parameters }
\label{SM:sect:ZFS}

\begin{table*}[h]
\setlength\extrarowheight{1pt}
\centering
\begin{tabular}{|l|l|cccc|}
\hline
State& Parameter &  MODEL-1 & MODEL-2 & MODEL-3 & MODEL-4  \\
\hline
\multirow{1}{*}{$(1)^3\bar{A}_2$} & $D$ (GHz) & 3.00 & 3.27 & 3.24 & 3.27 \\
\hline
\hline
\multirow{7}{*}{$(1)^3\bar{E}$} & $\lambda_z$ (GHz) & 38.3 & 19.9 & 21.5 & 18.7 \\ 
& $p$ (-)& 0.57 & 0.40 & 0.27 & 0.23 \\
& $p\lambda_z$ (GHz) & 21.8 & 8.02 & 5.70 & 4.32 \\
\cline{2-6}
& $D_{es}$ (GHz) & 1.89 & 2.18 & 2.11 & 2.11 \\
\cline{2-6}
& $\Delta$ (GHz) & 2.66 & 3.00 & 2.88 & 2.87 \\
& $q$ (-) & 0.79 & 0.70 & 0.63 & 0.62  \\
& $q\Delta$ (GHz)& 2.09 & 2.10 & 1.82 & 1.77  \\
\hline
\end{tabular}
\caption{Model-size dependence of zero-field splitting parameters. Ham reduction factors are defined as \\ $ p = e^{-1.974\left(\frac{\bar{E}_{\rm JT}(avg.)}{h\nu_e}\right)^{0.761}} $ and $q = (p + 1) / 2$. }
\label{SM:table:ZFS}  
\end{table*}